\documentclass{aa}

\bibpunct{(}{)}{;}{a}{}{,} 

\usepackage{graphicx}

\usepackage{longtable}
\usepackage{multicol}

\usepackage{savesym}
\usepackage{amsmath}
\savesymbol{iint}
\usepackage{graphicx}
\usepackage{xspace}
\usepackage{booktabs}
\usepackage[varg]{txfonts}
\usepackage{natbib,twoopt}
\usepackage[breaklinks=false, colorlinks=true, citecolor=blue]{hyperref}
\usepackage{epsfig}
\usepackage{txfonts}
\usepackage{amssymb}
\restoresymbol{TXF}{iint}

\newcommand{\parm}[1]{\ensuremath{#1}\xspace}

\newcommand{\pec}{\parm{e}}                               
\newcommand{\pk}{\parm{K}}                                
\newcommand{\pom}{\parm{\omega}}                  
\newcommand\psini{P_{rot} (\sin i_{\star})^{-1}}
\newcommand\ii{\rm I_{2}}

 \def\grcm2{\hbox{\,g\,cm$^{-2}$}}           
 \def\logg{$\log g$}                         
 \def\ms{\hbox{\,m\,s$^{-1}$}}               
 \def\m2s2{\hbox{\,m$^{2}$\,s$^{-2}$}}       
 \def\kms{\hbox{\,km\,s$^{-1}$}}             
 \def\gcm3{\hbox{\,g\,cm$^{-3}$}}            
 \def\vsini{\hbox{$v$\,sin\,$i$}}    
 \def\sini{\hbox{sin\,$i$}}          
 \def\Msun{\hbox{$M_{\odot}$}}               
 \def\Rsun{\hbox{$R_{\odot}$}}               
 \def\Mjup{\hbox{$\mathrm{M}_\mathrm{Jup}$}} 
 \def\redchi2{\parm{\chi_\mathrm{red}^{2}}}         

\usepackage{color}

\begin{document}

\title{Precise radial velocities of giant stars}
\subtitle{IX. HD~59686 Ab: a massive circumstellar planet orbiting a giant star in a $\sim$13.6~au eccentric binary system\thanks{Based on observations collected at the Lick Observatory, University of
  California.}\fnmsep\thanks{Based on observations collected at the Large Binocular Telescope, on Mount Graham, Arizona.}\fnmsep\thanks{Table 3 is only available in electronic form at the CDS via anonymous ftp to \href{http://cdsarc.u-strasbg.fr}{\color{blue}cdsarc.u-strasbg.fr (130.79.128.5)}  or via \href{http://cdsarc.u-strasbg.fr/viz-bin/qcat?J/A+A/595/A55}{\color{blue}http://cdsarc.u-strasbg.fr/viz-bin/qcat?J/A+A/595/A55}}}

\author{Mauricio~Ortiz\inst{1}
\and
Sabine~Reffert\inst{1}\and
Trifon~Trifonov\inst{1,2}\and
Andreas~Quirrenbach\inst{1}\and
David~S.~Mitchell\inst{3}\and
Grzegorz~Nowak\inst{10,11}\and
Esther~Buenzli\inst{4}\and
Neil~Zimmerman\inst{7,8}\and
Micka\"{e}l Bonnefoy\inst{9}\and
Andy~Skemer\inst{5}\and
Denis~Defr\`ere\inst{5}\and
Man~Hoi~Lee\inst{2,12}\and
Debra~A.~Fischer\inst{6}\and
Philip~M.~Hinz\inst{5}
}

\institute{Landessternwarte, Zentrum f\"{u}r Astronomie der Universität
  Heidelberg, K\"{o}nigstuhl 12, 69117 Heidelberg, Germany\\
  \email{mortiz@lsw.uni-heidelberg.de}   
  \and Department of Earth Sciences, The University of Hong Kong, Pokfulam           Road, Hong Kong                              
  \and Physics Department, California Polytechnic State University, San
  Luis Obispo, CA, 93407, USA
  \and Institute for Astronomy, ETH Zurich, Wolfgang-Pauli-Strasse 27, 8093 Zurich, Switzerland
  \and Steward Observatory, Department of Astronomy, University of Arizona, 933 N. Cherry Ave, Tucson, AZ 85721, USA
 \and Department of Astronomy, Yale University, New Haven, CT, 06511, USA
 \and Max-Planck-Institut f\"{u}r Astronomie, K\"{o}nigstuhl 17, 69117 Heidelberg, Germany
 \and Space Telescope Science Institute, 3700 San Martin Drive, Baltimore,
MD 21218, USA
 \and Univ. Grenoble Alpes, IPAG, F-38000 Grenoble, France. CNRS, IPAG, F-38000 Grenoble, France
 \and Instituto de Astrof\'isica de Canarias, 38205 La Laguna, Tenerife, Spain
 \and Departamento de Astrof\'isica, Universidad de La Laguna, 38206 La Laguna, Tenerife, Spain
 \and Department of Physics, The University of Hong Kong, Pokfulam Road, Hong Kong\\
}
\date{Received 26 April 2016 / Accepted 30 July 2016}

\abstract
{For over 12 yr, we have carried out a precise radial velocity (RV) survey of a sample of 373 G- and K-giant stars using the Hamilton \'Echelle Spectrograph at the Lick Observatory. There are, among others, a number of multiple planetary systems in our sample as well as several planetary candidates in stellar binaries.}
{We aim at detecting and characterizing substellar and stellar companions to the giant star HD~59686 A (HR~2877, HIP~36616).}
{We obtained high-precision RV measurements of the star HD~59686 A. By fitting a Keplerian model to the periodic changes in the RVs, we can assess the nature of  companions in the system. To distinguish between RV variations that are due to non-radial pulsation or stellar spots, we used infrared RVs taken with the CRIRES spectrograph at the Very Large Telescope. Additionally, to characterize the system in more detail, we obtained high-resolution images with LMIRCam at the Large Binocular Telescope.}
{We report the probable discovery of a giant planet with a mass of $m_{p}~\sini=6.92_{-0.24}^{+0.18}$~\Mjup~orbiting at $a_{p}=1.0860_{-0.0007}^{+0.0006}$~au from the giant star HD~59686 A. In addition to the planetary signal, we discovered an eccentric ($e_{B}=0.729_{-0.003}^{+0.004}$) binary companion with a mass of $m_{B}~\sini=0.5296_{-0.0008}^{+0.0011}$~\Msun~orbiting at a close separation from the giant primary with a semi-major axis of $a_{B}=13.56_{-0.14}^{+0.18}$~au.}
{The existence of the planet HD~59686~Ab in a tight eccentric binary system severely challenges standard giant planet formation theories and requires substantial improvements to such theories in tight binaries. Otherwise, alternative planet formation scenarios such as second-generation planets or dynamical interactions in an early phase of the system's lifetime need to be seriously considered to better understand the origin of this enigmatic planet.}

\keywords{Techniques: radial velocities -- Planets and satellites: detection --
  Individual: (HD~59686 HIP~36616 HR~2877) -- Giant planets -- Binaries: spectroscopic}

\maketitle

\section{Introduction}
Since the discovery of the first extrasolar planet around a solar-like star 20 yr ago by \citet{Mayor1995}, more than 3000 exoplanets have been confirmed \citep{Schneider2011}\footnote{\href{http://www.exoplanet.eu}{\color{blue}http://www.exoplanet.eu}}. Of these planets, about 75\%\ have been discovered by the transit method, which has greatly benefited from the data obtained with the \textit{Kepler} Space Telescope \citep{Borucki2010}. Moreover, in 2014 alone, 715 new planets in 305 systems were detected by \textit{Kepler}, almost doubling the number of exoplanets known at that time \citep{Lissauer2014,Rowe2014}, and more recently, \citet{Morton2016} have confirmed nearly 1280 new transiting \textit{Kepler} planets based on probabilistic validation methods. Most of the remaining 25\%\ of planets have been found using the radial velocity (RV) technique. The spectral characteristics
of solar-type main-sequence (MS) stars are favorable  for RV measurements, which has made these stars the targets of the majority of the RV planet searches. However, a growing number of research
groups are successfully searching for planets around evolved subgiant and giant stars
\citep[e.g.,][]{Dollinger2009,Johnson2010,Johnson2011,Sato2013,Jones2015a,
Jones2015b,Niedzielski2015,Reffert2015,Wittenmyer2016}.

There are currently 95 known planetary companions orbiting around giant
stars, of which 46\% have been published during the past three years\footnote{\href{http://www.lsw.uni-heidelberg.de/users/sreffert/giantplanets.html}{\color{blue}http://www.lsw.uni-heidelberg.de/users/sreffert/giantplanets.html}}. Giant stars allow us to access more massive stars than those typically observed on the MS. Early MS stars are normally avoided in RV planet searches as they rotate faster and have too few absorption lines for reliable high-precision RV determinations. On the other hand, evolved stars, such as K~giants, have suitable and less broadened absorption lines for RV measurements, low rotational velocities, and much higher masses than late-type MS stars. Additionally, K-giant RV surveys also allow investigating how planetary systems evolve after the host star leaves the MS \citep{Villaver2009,Kunitomo2011,Villaver2014}.

Of the known extrasolar planets, $\text{about }7\%$ orbit in multiple star systems\footnote{\href{http://www.univie.ac.at/adg/schwarz/multiple.html}{\color{blue}http://www.univie.ac.at/adg/schwarz/multiple.html}}, although this number suffers from an observational bias as most of the exoplanet surveys systematically avoid binary stars in their samples. For K-giant stars specifically, only four out of 72 known stars harboring planets are members of stellar multiple systems: 11 Com \citep{Liu2008}, $\gamma^{1}$ Leo \citep{Han2010}, 91 Aqr \citep{Mitchell2013}, and 8 UMi \citep{Lee2015}. Finding planets in multiple star systems allows us to learn more about the processes of planetary formation and evolution. This is particularly important, since $\sim50$\%\ of the MS stars in our solar neighborhood are members of binaries or multiple systems \citep{Duquennoy1991,Raghavan2010}. The frequency of these planets may have a strong influence on the overall global frequency of extrasolar planets, allowing us to study the efficiency of planet formation mechanisms. Moreover, if there is any difference in the properties of planets in binary systems with respect to planets orbiting single stars, this may unveil effects caused by additional companions in stellar systems \citep{Desidera2010, Roell2012}.

The majority of known planets in binary systems are in S-type orbits (circumstellar planets), meaning that the planet orbits around one member of the binary pair \citep[e.g.,][]{Howard2010,Buchhave2011,Anderson2014}, as opposed to P-type configurations (circumbinary planets), where the planet orbits both stars beyond the binary orbit \citep[e.g.,][]{Doyle2011,Orosz2012a,Orosz2012b,Welsh2012,Bailey2014}. Most of the known S-type planets reside in wide-separation binaries ($a_{B}\gtrsim100$~au) where the influence of the stellar companion on the formation process of the inner planet can probably be neglected. However, there are some interesting systems detected in close-separation binaries in which the stellar companion is located at roughly $20$~au: Gliese 86  \citep{Queloz2000}, $\gamma$ Cep \citep{Hatzes2003}, HD~41004 \citep{Zucker2004}, and HD~196885 \citep{Correia2008}. The existence of planets in tight binary systems ($a_{B}\lesssim20$~au) presents a serious challenge to current planet formation theories \citep{Hatzes2005,Rafikov2005}. Moreover, supporting the theoretical expectation, \citet{Wang2014} found evidence that planet formation is effectively suppressed in binary systems with separations of less than 20 au. 

In this work, we report the discovery of a planet orbiting the giant star HD~59686, which we refer to as HD~59686~A, and which is part of a close-separation binary system with $a_{B}=13.56$~au. The paper is organized as follows: In Sect.~2 we describe our sample selection and observations.  Section~3 presents the stellar properties of the star and the Keplerian fit to the RV data from the Lick observatory. In Sect. 4 we validate the planetary hypothesis for the RV variations in HD~59686 A using infrared RVs taken with CRIRES, spectral activity indicators, and the available photometry. In Sect.~5 we describe the high-contrast imaging observations of HD~59686 A obtained with LMIRCam at the Large
Binocular Telescope (LBT) to image the stellar companion, including reduction of the data and constraints on the stellar companion to the giant star.  In Sect.~6 we discuss the properties of the HD~59686 system, focusing on the nature of the stellar companion and the implications for the formation of planets in tight binaries. Finally, in Sect.~7 we present our conclusions.

\section{Observations}
We have continuously monitored the RVs of  373 G- and K-giant stars for more than a decade, resulting in several published planet detections \citep{Frink2002,Reffert2006,Quirrenbach2011,Mitchell2013,Trifonov2014}. Typical masses in our sample are between $\sim$1--3~\Msun~and we reached RV precisions of $\sim$5--8 \ms. Among other results, we have found the first planet around a giant star \citep{Frink2002} and showed that red giants with masses greater than $\sim$2.7~\Msun\ host very few giant planets with an occurrence rate lower than 1.6\% \citep{Reffert2015}. 

The original selection criteria aimed at observing 86 bright K-giant stars ($V\leqslant6$~mag) that were not variable or part of multiple stellar systems. Later in the survey, 93 new stars were added to the sample by imposing less stringent constraints regarding the photometric stability. Finally, in 2004, we added 194 G and K giants with bluer colors ($0.8\leqslant B-V\leqslant1.2$) with the aim of reducing the intrinsic RV jitter  \citep[e.g.,][]{Frink2001, Hekker2006}. The inclusion of these stars allowed us to probe higher masses to test whether more massive stars host more massive planetary companions. More details on the selection criteria and on the giant star sample can be found in \citet{Frink2001} and  \citet{Reffert2015}.

The RV observations of HD~59686 A were carried out using the Hamilton \'Echelle Spectrograph \citep{Vogt1987} fed by the 0.6~m Coud\'{e} Auxiliary Telescope (CAT) of the Lick Observatory (California, USA). The Hamilton spectrograph covers the wavelength range 3755--9590~{\AA} and has a resolution of $R\,{\sim}\,60\,000$. The data were acquired and reduced using the iodine cell approach as described by \citet{Butler1996}. We currently have 11--12 yr of data for our original set of K-giant stars,
of which HD~59686 A is a member. In total, we have 88 RV measurements for HD~59686~A throughout this period of time.  The Lick RVs along with their formal uncertainties are listed in Table 3. Typical exposure times were 20~min, and  the signal-to-noise ratios for these observations are typically about 120--150. The resulting RV measurements have a median precision of $\sim$5.6~\ms. This value is below the RV jitter of $16.4\pm2.9$~\ms~expected for HD~59686 A based on scaling relations \citep[see][]{Chaplin2009,Kjeldsen2011}. Additionally, using our K-giant sample, we have obtained an empirical relation for the expected RV jitter as a function of color \citep[see][]{Frink2001,Trifonov2014} given by

\begin{equation}
\log (\textrm{RV jitter [m/s]}) = (1.3\pm0.1)(B-V) + (-0.04\pm0.1)
\label{jitter}
,\end{equation} 

\noindent where $(B-V)$ is the color index. Using this relation, we expect an intrinsic RV jitter of $26.5\pm9.2$ \ms~for HD~59686~A. This value is consistent at the 1.1$\sigma$ level with the result derived from scaling relations.


\begin{table}
\caption{Stellar parameters of HD~59686 A.}
\label{param}
\centering
\begin{tabular}{lc}
\hline
\hline
\noalign{\smallskip}
Parameter & Value\\
\noalign{\smallskip}
\hline
\noalign{\smallskip}
Apparent magnitude $m_{\rm{v}}$ (mag)\tablefootmark{a} & 5.45\\

Absolute magnitude $M_{\rm{v}}$ (mag)                            & $0.52 \pm 0.06$\\

Near-infrared magnitude $K$ (mag)\tablefootmark{b}                & $2.92 \pm 0.30$\\

Color index $B-V$  (mag)\tablefootmark{a}          & $1.126 \pm 0.006$\\

Effective temperature $T_{\rm{eff}}$ (K)\tablefootmark{c}   & $4658 \pm 24$\\

Surface gravity $\log g\ {\rm (cm~s^{-2})}$\tablefootmark{c} & $2.49 \pm 0.05$\\

Metallicity [Fe/H] (dex)\tablefootmark{d}                     & $0.15 \pm 0.1$\\

Stellar mass $M_\star$ (\Msun)\tablefootmark{c}       & $1.9 \pm 0.2$\\

Stellar radius $R_\star$ (\Rsun)\tablefootmark{c}        & $13.2 \pm 0.3$\\

Parallax (mas)\tablefootmark{a}           & $10.33 \pm 0.28$\\

Distance (pc)                                 & $96.8 \pm~^{2.7}_{2.6}$\\

Age (Gyr)\tablefootmark{c}                & $1.73 \pm 0.47$\\

Spectral type                                       & K2 III\\

\noalign{\smallskip}
\hline
\end{tabular}
\tablefoot{\\
\tablefoottext{a}{Data from Hipparcos:~\citet{vanLeeuwen2007}}\\
\tablefoottext{b}{Data from 2MASS:~\citet{Skrutskie2006}}\\
\tablefoottext{c}{\citet{Reffert2015}}\\
\tablefoottext{d}{\citet{Hekker2007}}
}
\end{table}

\section{Results}
\subsection{Stellar properties}
The stellar properties of the giant star HD~59686 A are given in Table~\ref{param}. HD~59686 A is a slightly metal-rich star with [Fe/H]=$0.15\pm0.1$~dex \citep{Hekker2007}. To derive the stellar mass, we interpolated between the evolutionary tracks \citep{Girardi2000}, stellar isochrones, and metallicities using a trilinear interpolation scheme. This approach commonly allows two possible solutions depending on the evolutionary status of the star, namely red giant branch (RGB) or horizontal branch (HB). By taking the evolutionary timescale, that is, the speed with which the star moves through that portion of the evolutionary track, as well as the initial mass function into account, probabilities were assigned to each solution. The derived mass of HD~59686 A is $M_\star=1.9\pm0.2$~\Msun~and the star was found to have a 89\%\ probability of being on the HB.  If it were instead on the RGB, then it would have a mass of $2.0 \pm 0.2$~\Msun, thus the mass is not affected within the uncertainties. More details on the method, including the stellar parameters of all K-giant stars in our Doppler survey, can be found in \citet{Reffert2015}.

The angular diameter of HD~59686 A was first calculated by \citet{Merand2004}, using absolute spectro-photometric calibration from IRAS and 2MASS observations. They derived a diameter of $1.29 \pm 0.02$~mas, which at the Hipparcos distance of $96.8^{+2.7}_{-2.6}$~pc gives a value for the radius of $13.4\pm0.4$~\Rsun. Later, \citet{Baines2008a} used the Center for High Angular Resolution Astronomy (CHARA) interferometric array \citep{Brummelaar2005} to measure an angular diamater of $1.106\pm0.011$~mas, deriving a radius of $11.62\pm0.34$~\Rsun. Our estimate of the stellar radius for HD~59686 A of $R_\star=13.2\pm0.3$~\Rsun~ agrees well with the value derived by \citet{Merand2004} and  is slightly higher than the one obtained by \citet{Baines2008a}.

\subsection{Keplerian orbits}
We fitted Keplerian orbits to the RV data of HD~59686 A. The uncertainties were derived through bootstrapping (using 5\,000 bootstrap replicates) by drawing synthetic samples from the original RV dataset with data replacement \citep[see][]{Press1992}. We fitted for two companions in the system, to which we refer as HD~59686 Ab and HD~59686~B.


\begin{table}
\caption{Orbital parameters of the HD~59686 system.}
\label{hip36616orbit}
\centering
\begin{tabular}{lcc}
\hline\hline
\noalign{\smallskip}
Parameter & HD~59686 Ab & HD~59686 B \\
\hline
\noalign{\smallskip}
$P$ (days) & $299.36_{-0.31}^{+0.26}$ & $11680_{-173}^{+234}$\\
\noalign{\smallskip}
$M_{0}$ (deg)\tablefootmark{a} & $301_{-85}^{+26}$ & $259_{-1}^{+3}$\\
\noalign{\smallskip}
$e$ & $0.05_{-0.02}^{+0.03}$ & $0.729_{-0.003}^{+0.004}$\\
\noalign{\smallskip}
$\omega$ (deg) & $121_{-24}^{+28}$ & $149.4_{-0.2}^{+0.2}$\\
\noalign{\smallskip}
$K$ (\ms) & $136.9_{-4.6}^{+3.8}$ & $4014_{-8}^{+10}$\\
\noalign{\smallskip}
\hline
\noalign{\smallskip}
$m$~\sini~(\Mjup) & $6.92_{-0.24}^{+0.18}$ & $554.9_{-0.9}^{+1.2}$\\
\noalign{\smallskip}
$a$ (au) & $1.0860_{-0.0007}^{+0.0006}$ & $13.56_{-0.14}^{+0.18}$\\
\noalign{\smallskip}
\hline
\end{tabular}
\tablefoot{\\
\tablefoottext{a}{This parameter is the value of the mean anomaly at the first observational epoch $t_{0}=2451482.024$~JD.}
}
\end{table}

\begin{figure*}[]
\resizebox{\hsize}{!}{\includegraphics{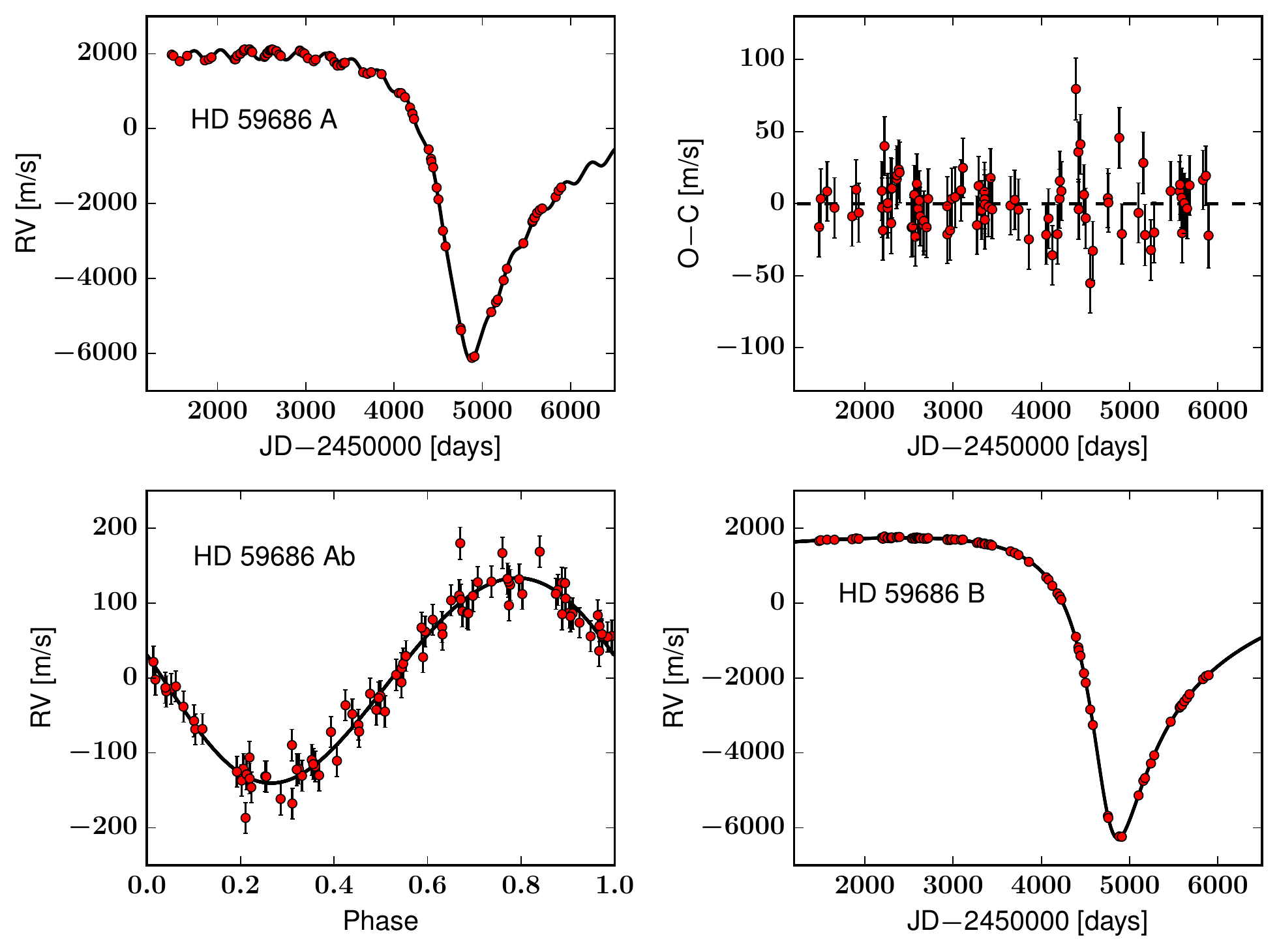}}
\caption{ Radial velocity measurements of the HD~59686 system. Note that a jitter of 19.83 \ms~was added in quadrature to the formal RV uncertainties, and this is reflected in the plot. \textit{Upper left}: Lick RVs together with the best Keplerian fit to the data. \textit{Upper right}: RV residuals from the fit. \textit{Lower left}: Phased RV variations and Keplerian fit for the $\sim7$~\Mjup~planet HD~59686 Ab after subtracting the signal of the stellar companion. \textit{Lower right}: RV data and Keplerian fit for the $\sim0.5$~\Msun~stellar object HD~59686 B after subtracting the planetary signal.}
\label{hip36616rv}
\end{figure*}

In total, the Keplerian fit for HD~59686~Ab and HD~59686~B has 11 free parameters: the orbital period $P$, argument of periastron \pom, RV semi-amplitude \pk, mean anomaly $M_{0}$, and eccentricity \pec for each of the companions, and an arbitrary zero-point offset. The RVs of HD~59686~A are shown in Fig.~\ref{hip36616rv} along with the best Keplerian fit to the data. We also plot the individual signals of the planet HD~59686~Ab and the stellar object HD~59686~B. Error bars are included in all the plots. 

The best-fit orbital parameters for the planetary and stellar companions are given in Table~\ref{hip36616orbit}. It is worth mentioning that K-giant stars exhibit intrinsic RV variability, known as stellar jitter. Therefore we decided to add in quadrature a jitter of 19.83 \ms~, coming from the rms of the residuals around the fit, to our formal RV errors, which scaled down the \redchi2~to a value of 1 (without jitter, \redchi2=11.7). The rms of the residual RVs, after subtracting the best Keplerian fit that includes the jitter, is 19.49 \ms. This result is consistent with the intrinsic scatter expected from K-giant stars \citep[Eq. \ref{jitter}, see also][]{Hekker2008} and is within 1.1$\sigma$~from the value derived using scaling relations.

\begin{figure}[]
\resizebox{\hsize}{!}{\includegraphics{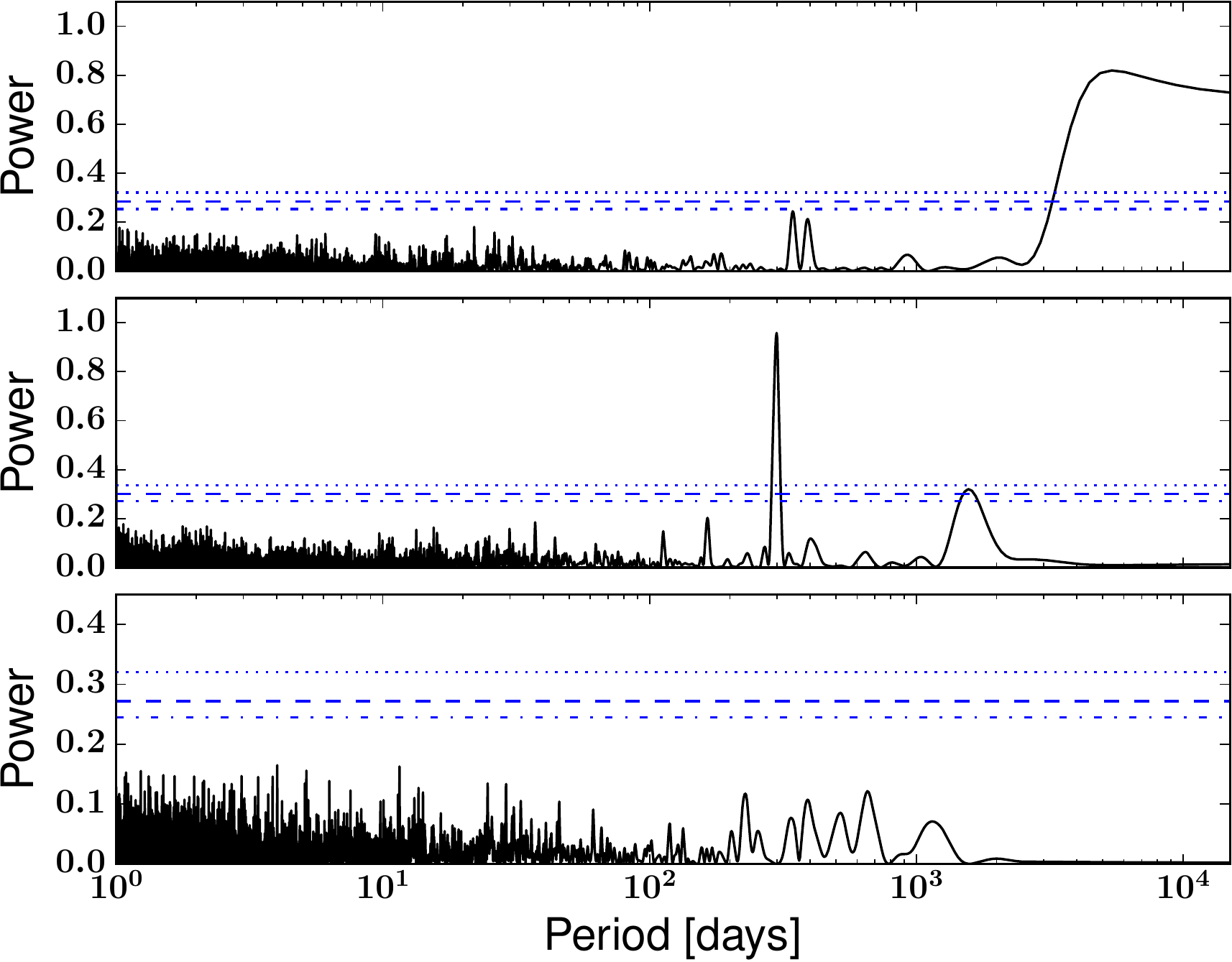}}
\caption{\textit{Top}: GLS periodogram of the RV data of HD~59686 A.  The significant peaks at $\sim$5000 and $\sim$340~days represent the orbits of the stellar and planetary companions, respectively. The 5000-day period is the time frame of our observations, therefore it is much shorter than the actual stellar period. The dotted, dashed, and dash-dotted lines show FAPs of 0.1\%, 1\%, and 5\%, respectively. \textit{Middle}: Periodogram of the residual RVs after the signal due to the stellar companion is removed from the data. Now the peak due to the planetary companion becomes much stronger and the alias period at $\sim$400~days disappears. Additionaly, another alias period appears at $\sim$1700~days. \textit{Bottom}: Periodogram of the residual RVs after subtracting the orbital fit (stellar+planetary companions); it shows no significant peaks.}\label{hip36616losca}
\end{figure}

Figure~\ref{hip36616losca} shows a generalized Lomb-Scargle (GLS) periodogram \citep{Zechmeister2009} of HD~59686 A RVs.  The top panel shows the results for the RV data, while the middle and bottom panels show the periodogram for the residuals after subtracting the stellar and stellar+planetary signals, respectively. The false-alarm probabilities (FAPs) were calculated by replacing the original RVs with randomly scrambled data through bootstraping. We computed the GLS periodogram 1\,000 times for this new dataset and calculated how often a certain power level is exceeded just by chance. The estimated FAPs of 0.1\%, 1\%, and 5\% are shown in the plot as the horizontal dotted, dashed, and dash-dotted blue lines, respectively. 

The top panel shows one significant peak in the GLS periodogram at $\sim$5000~days, which is approximately the length of time over which HD~59686~A has been observed.  This wide peak represents the long period of HD~59686~B ($P=11680_{-173}^{+234}$~days), for which one complete orbit has not been observed yet. However, we are able to set tight constraints on the binary period as the eccentricty of the orbit is very high. The second strongest peak is at $\sim$340~days, very roughly matching the best Keplerian fit for the planetary companion ($P=299.36_{-0.31}^{+0.26}$~days). The third largest peak at $\sim$400~days is an alias period that disappears when the signal of the stellar companion is removed from the data. This is shown in the middle panel where the strength of the signal due to the planet increases significantly, and another alias period also appears at around 1700 days. After the planetary companion is subtracted, this peak disappears and no significant periodicities are observed in the signal of the RV residuals.

By adopting a stellar mass of $M_\star=1.9\pm0.2$~\Msun, we derived a minimum mass of $6.92_{-0.24}^{+0.18}$~\Mjup~for HD~59686 Ab and a value of $554.9_{-0.9}^{+1.2}$~\Mjup~for HD~59686 B. The mass for HD~59686 B is equivalent to $\sim0.53$~\Msun, which immediately places this companion in the stellar regime; it cannot be a massive planet or a brown dwarf. The planet orbits the giant star at a distance of $a_{p}=1.0860_{-0.0007}^{+0.0006}$~au, while the semi-major axis of the stellar companion is $a_{B}=13.56_{-0.14}^{+0.18}$~au. Furthermore, the orbit of HD~59686 B is very eccentric ($e=0.729_{-0.003}^{+0.004}$), which may have played an important role in the formation and/or evolution of the inner planet, as we discuss in Sect. 6. 


\section{Validating the planetary signal}
\subsection{Rotational modulation}
Stellar surface phenomena such as star spots, plages, or convection modulated by stellar rotation may generate low-amplitude RV variations that can mimic planetary signatures. To investigate such false-positive scenarios, we determined the stellar rotation of HD~59686 A.  \citet{Hekker2007} estimated the projected rotational velocity of HD~59686 A to be $\vsini=4.28\pm1.15$~\kms. Using our estimate of the stellar radius ($R_\star$ = 13.2 $\pm$ 0.3 $\Rsun$), we determine an upper limit for the rotation period of HD~59686 A of $\psini$ = 156.03 $\pm$ 39.35 days. This means that any low-amplitude RV variations generated by surface phenomena that are modulated by stellar rotation cannot have periods longer than $\sim$195 days. Therefore, it is unlikely that the periodic signal ($P = 299.36_{-0.31}^{+0.26}$~days) is generated by stellar rotation.

\citet{Massarotti2008} estimated a slightly lower value for the projected rotational velocity of $3.8$~\kms~for HD~59686~A, which implies $\psini$ = 175.74 $\pm$ 46.42 days (assuming 1 \kms~error in \vsini), consistent with the results of \citet{Hekker2007} and with the above statement. On the other hand, \citet{Carlberg2012} calculated a value of $\vsini=0.93\pm0.41$~\kms~for the K-giant, implying $\psini$ = 718 $\pm$ 495 days. However, this result has large uncertainties and, as discussed by the authors, their estimates of $\vsini$ show significant systematic differences when compared to values derived in the literature. Specifically, their estimates of \vsini~are systematically lower than those reported in other studies \citep[see Fig. 6 of][]{Carlberg2012}, which can be accounted for by an overestimation of the macroturbulence velocity, particularly in the slow-rotation regime.

Regardless of the above considerations, to test the hypothesis that the 299.36$_{-0.31}^{+0.26}$ day period may be caused by stellar activity, like long-period pulsations for example, we checked infrared RVs, all available photometry, and spectral activity indicators as described in the following sections.

\subsection{Infrared radial velocities}
It is recognized that intrinsic stellar activity, such as cool spots, can create RV variations in giant stars that can mimic the presence of companions \citep[e.g.,][]{Hatzes2000,Hatzes2004}. This poses an additional challenge for validating the interpretation of a periodic RV change as a \textit{\textup{bona fide}} planet, when compared to inactive MS stars. Moreover, some giant stars are known to be pulsating stars, which show several modes of pulsation with varying amplitudes and frequencies \citep{DeRidder2009,Huber2010,Christensen2012,Stello2013}. In the case of radial pulsations, the stellar surface moves away and toward the observer, which induces periodic RV variations.

The pulsation frequencies of a star are closely related to its density
and temperature, as these control the speed at which sound waves can
propagate.  Using the scaling relation of \citet{Kjeldsen1995}, we calculated the period of the pulsation with maximum amplitude using our derived values of the radius, mass, and effective temperature,
which yielded a value of 0.31 days for HD~59686 A. Although this calculation is not ideal for giant stars, it should give a reasonable estimate of the pulsation period with the largest amplitude. This value is orders of magnitudes below the RV oscillations seen in our data.


\begin{figure}
\resizebox{\hsize}{!}{\includegraphics{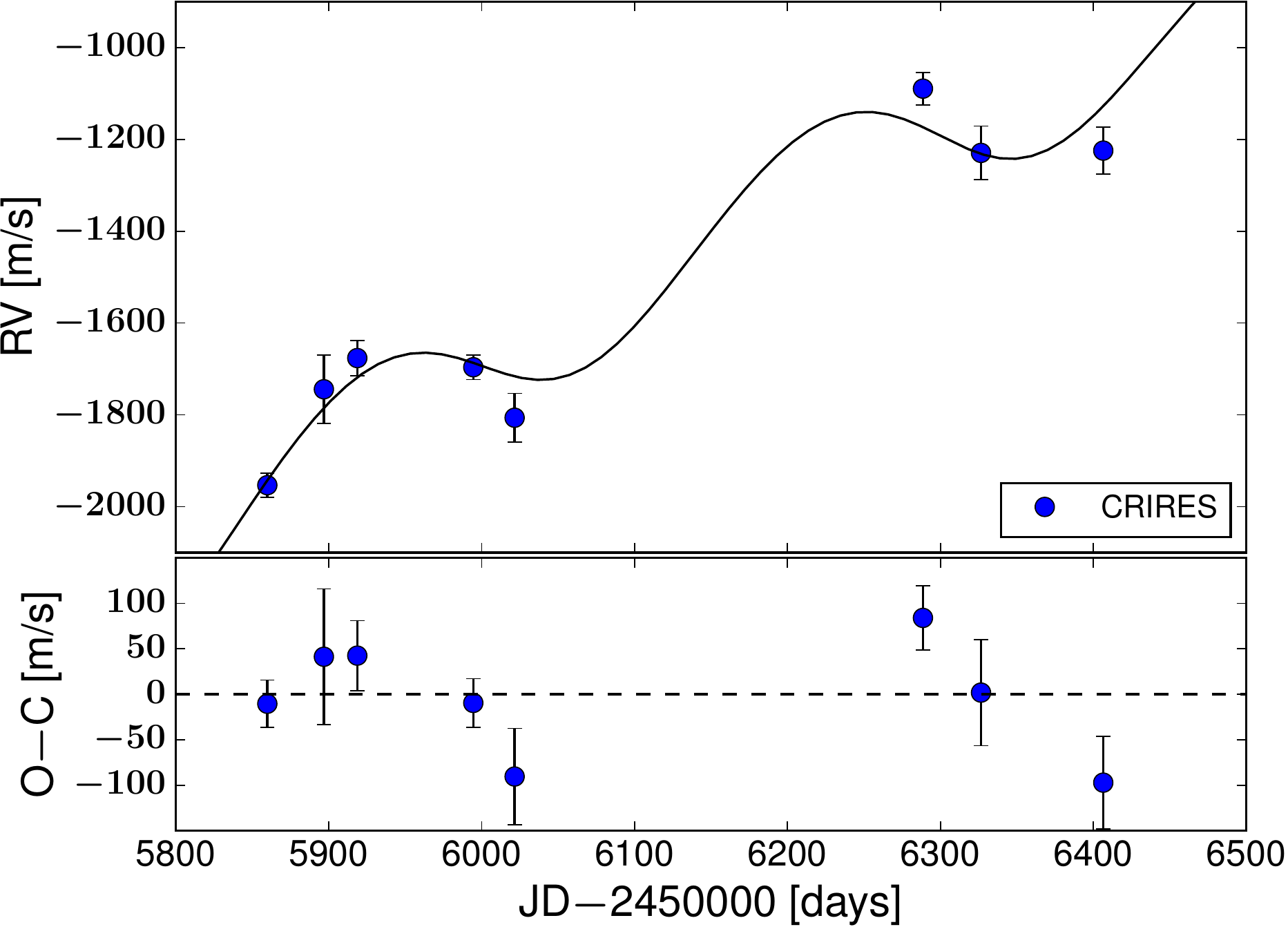}}
\caption{\textit{Upper panel}: CRIRES infrared RV measurements of HD~59686~A. The black solid line shows the best Keplerian fit obtained from the Lick data alone. \textit{Bottom panel}: Residuals of the CRIRES RVs from the optical fit. The value of the rms is $\sim$59~\ms, which is consistent with the large infrared RV errors. The mean error of the CRIRES data, with a jitter of 19.83 \ms~added in quadrature,  is $\sim$45~\ms.}
\label{36616crires}
\end{figure}

It is possible, though unlikely, that some pulsation exists in HD~59686 A with
a much lower frequency, but large enough amplitude to be detectable in our
data, which could be the source of the RV oscillations we observe. Non-radial pulsations are much more complicated to model, and they can display an arbitrary number of amplitudes and periods for different modes. However, it is not expected that the RV amplitude of the pulsations in the visible waveband match the amplitude in the infrared, since the photometric variations of pulsating stars are wavelength dependent \citep[e.g.,][]{Percy2001,Percy2008}.  On the other hand, if the RV oscillations are due to a companion, then the infrared and visible RV variations should be consistent with each other. 

In 2012 and 2013, \citet{Trifonov2015} obtained infrared RVs of HD~59686 A using the CRyogenic high-resolution InfraRed Echelle Spectrograph \citep[CRIRES;][]{Kaufl2004} at the Very Large Telescope (VLT), in Chile. Their CRIRES spectra have a resolution of $R\,{\sim}\,100\,000$ and cover the wavelength range 1.57--1.61~$\mu$m. Details of the CRIRES observations and the reduction process, including the measured RVs for HD~59686 A, can be found in \citet{Trifonov2015}.

We obtained the RV offset between the CRIRES and Lick velocities for HD~59686 A by fitting the CRIRES and Lick RVs keeping all the orbital parameters fixed. Figure~\ref{36616crires} shows the CRIRES RV data (with the RV offset applied) together with the best Keplerian fit to the Lick data. The infrared RVs match the Keplerian model obtained from the optical data. This should in general not be the case if the RV variations were due to large amplitude stellar pulsations. Moreover, the scatter around the fit of $\sim$59~\ms~is consistent with the relatively large uncertainties\footnote{To be consistent with the optical fit, a jitter of 19.83 \ms~was added to the formal CRIRES RV uncertainties.} of the CRIRES RVs that are on the order of $\sim$45~\ms. 

An additional test can be made by fitting only the CRIRES data to derive the RV semi-amplitude, $K_{IR}$. Following \citet{Trifonov2015}, we first subtracted the signal of the stellar companion from the CRIRES data. As the presence of HD~59686 B is clearly detected in the system, it is fair to assume that the RV signal due to this star is consistent in the optical and infrared data sets. Then, we performed a Keplerian fit to the CRIRES RVs keeping all parameters fixed (the parameters obtained from the Lick RVs) with the exception of the RV semi-amplitude and RV zero point. We derived a value of $K_{IR}=206.0\pm29.1$~\ms. The RV semi-amplitude of $K_{opt}=136.9_{-4.6}^{+3.8}$~\ms~from the optical RVs is within 2.25$\sigma$ from the IR value. If we calculate $\kappa=K_{IR}/K_{opt}$  , as in \citet{Trifonov2015}, then we obtain a value of $\kappa=1.50\pm0.22$, but we note that the calculated error might be underestimated as the error on the fitting of the stellar component is not taken into account. This result shows that the near-IR signal is not flat or of a smaller amplitude than the optical one, which we would expect for a spot or pulsations; the amplitude of pulsations decreases with increasing wavelength in pulsating giant stars \citep{Huber2003,Percy2008}. We also have only a handful of moderately precise IR RVs and in addition, a stellar jitter of about 20 \ms~for HD 59686~A, but we observe that the optical and near-IR phases are consistent, which is not necessarily expected for pulsations. This means that most likely the signal is real and caused by the gravitational perturbation of a companion in the system.

\begin{figure}
\resizebox{\hsize}{!}{\includegraphics{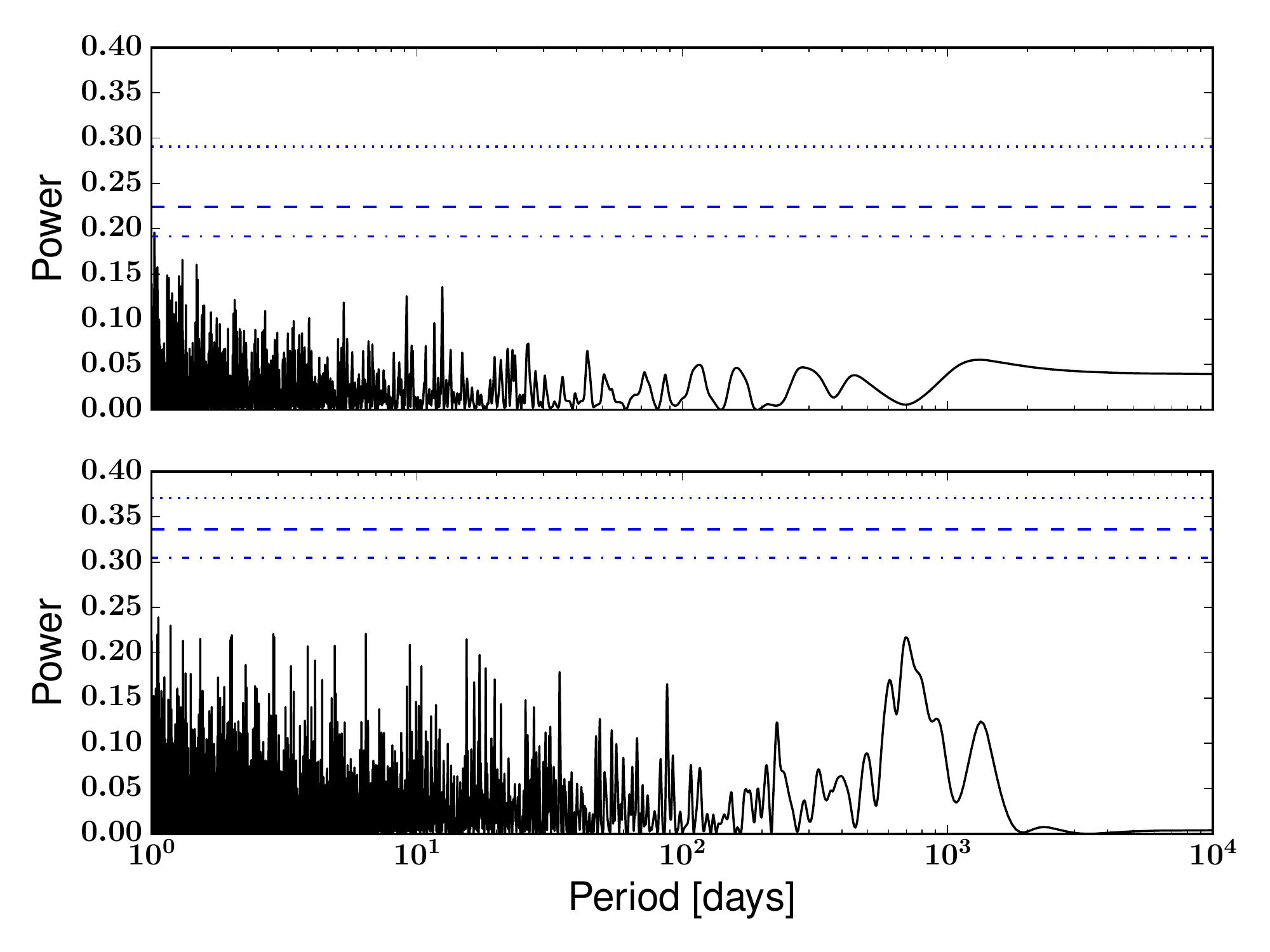}}
\caption{\textit{Upper panel}: GLS periodogram of the H${\alpha}$ index measurements of HD~59686 A. \textit{Bottom panel}: GLS periodogram of the Hipparcos V-band photometry of HD~59686 A. The dotted, dashed, and dash-dotted lines in both panels show FAPs of 0.1\%, 1\%, and 5\%, respectively, obtained by bootstraping. No significant periodicities are found in the data.}
\label{index_phot_period}
\end{figure}

\subsection{Photometry}
The ASAS-3 Photometric V-band Catalog \citep{Pojmanski1997,Pojmanski2001} contains 290 best-quality measurements (grade A) of HD~59686 A collected over seven years between December 13, 2002 (HJD = 2452621.84) and November 24, 2009 (HJD = 2455159.78). Unfortunately, HD~59686 A is a very bright target (V=5.45) and exceeds the ASAS-3 V-band saturation limit with the used exposure times (180 seconds). The high dispersion of the ASAS-3 V-band photometric measurements of HD~59686 A (peak-to-peak amplitude of 0.784 mag, mean value $\bar{V} = 5.74 \pm 0.19$ mag) and the mean value of the errors (38.5 mmag) ensure that HD~59686 A saturates the ASAS-3 detector.

The only unsaturated photometry for HD~59686 A was acquired by the Hipparcos mission \citep{ESA1997} between March 16,$^{}$ 1990 and March 10,$^{}$ 1993 (2447966.9 -- 2449057.2 JD), more than six years before first RV observations of HD~59686~A. The Hipparcos data set consists of 96 measurements with 5.6 mmag mean error, 5.6 mag mean value, and a standard deviation of 5.5 mmag, similar to the mean error of the measurements. As shown in the bottom panel of Fig. \ref{index_phot_period}, no significant periodic signal was found in the photometry of these data. Additionally, we can use the Hipparcos data to investigate whether a hypothetical spot might have produced a noticeable photometric variation. We derived the spot filling factor that would be required to generate the observed RV amplitude of $\sim$137 \ms~using the relation found by \citet{Hatzes2002} for cool spots on sun-like stars. We obtained a spot filling factor of $f=0.1$, meaning that 10\% of the stellar surface must have been covered by spots to produce the large RV variation seen in the data. Using this value for the filling factor, the expected photometric variability is $ \Delta m=0.078$~mag for a temperature difference of $\Delta T=1200$~K between the spot and the stellar photosphere. This level of variation is one order of magnitude above the observed dispersion seen in the Hipparcos data. The same is true for a wide range of temperature differences of typical star spots ranging from $\Delta T=200-1200$ K \citep[e.g.,][]{Biazzo2006,Oneal2006}.

Nevertheless, hypothetic surface structure phenomena might mimic the presence of an exoplanet. For example, \citet{Hatzes2000} investigated the possible existence of a macroturbulent spot to explain the RV variation of Polaris. Given the right conditions, this dark spot might cause a large RV oscillation without a significant photometric variation. However, the values of, for example, the magnetic field and the difference between the velocity of the macroturbulent spot and the surrounding surface must be exceptionally well fine-tuned to produce an RV variation of hundreds of m/s. In addition, if a macroturbulent spot causes the RV changes in HD 59686 A, then it must have been long-lived and maintained a constant and consistent effect during at least 12 yr. The same is true for long-lived long-period non-radial pulsations, which is not necessarily expected. Thus, although we cannot completely discard this scenario, a giant planet orbiting the star HD 59686 A appears as the most plausible interpretation of our data.

\begin{figure}
\resizebox{\hsize}{!}{\includegraphics{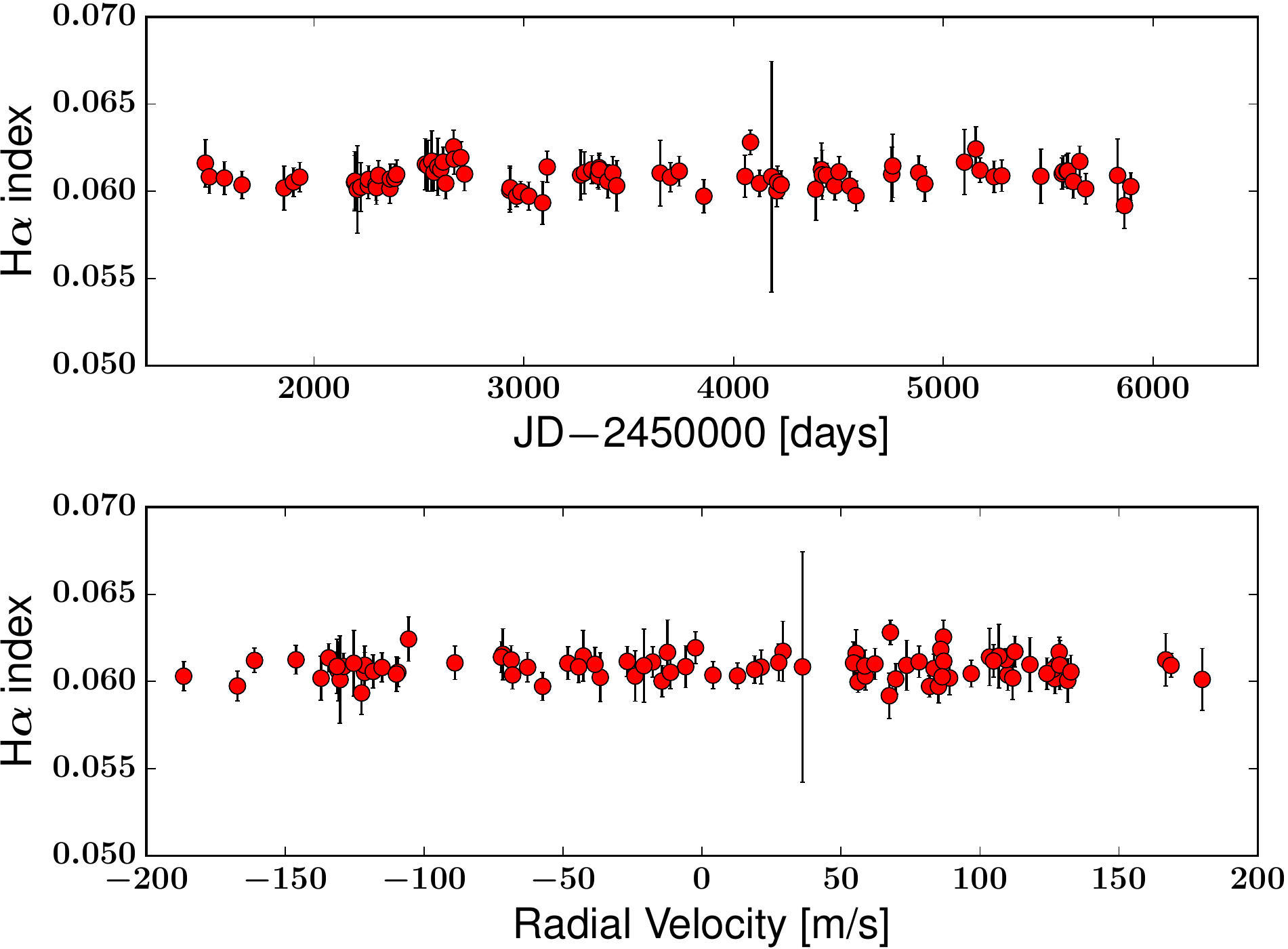}}
\caption{\textit{Upper panel}: H$\alpha$ index measurements at the time of each RV observation of HD~59686 A. \textit{Bottom panel}: H$\alpha$ index measurements as a function of the RVs due to the planetary companion HD~59686~Ab, that is, with the stellar component subtracted from the data. No significant correlation is seen in the data, which corroborates that a giant planet is part of the system.}
\label{index_time}
\end{figure}

\subsection{Spectral activity indicators}
Since the RV measurements of HD~59686~A were acquired using the iodine-cell method, it is difficult to perform precise bisector measurements of spectral lines as the stellar spectra are affected by $\ii$ lines. Instead of this, we performed an analysis of the H$\alpha$ line, which is located in the central region of one of the Hamilton spectrograph orders and is known to be a good indicator of stellar activity. We measured the H$\alpha$ index using the approach presented by \citet{Kurster2003}. However, we broadened the width of the window centered on H$\alpha,$ from $\pm$15~\kms~used by \citet{Kurster2003} for Barnard's star to $\pm$45.68~\kms~($\pm$1~{\AA}) recently used by \citet{Hatzes2015} for Aldebaran. As reference windows we used spectral regions that extend from $-250$ and $-650$~\kms~and from 250 and 650~\kms. The upper panel of Fig.~\ref{index_phot_period} shows the GLS periodogram of the H$\alpha$ index measurements. As for the Hipparcos photometry, no significant signal exist in the H$\alpha$ index of HD~59686~A. Figure \ref{index_time} presents the H$\alpha$ index against the time of each RV observation of HD~59686~A and as a function of the RV variation induced by the planet HD~59686~Ab (without the contribution of the stellar companion). The plot
shows no correlation between these RVs and the H$\alpha$ index. Moreover, we measured a Pearson correlation coefficient of $r$ = 0.06 with a $p$-value=0.58. This analysis corroborates that the 299.36$_{-0.31}^{+0.26}$ day period in the RV curve of HD~59686~A is most likely generated by the gravitational pull of a planetary companion.

It is worth to note, however, that HD 59686 A shows some similarities to carbon-enhanced metal-poor (CEMP) stars \citep[see][]{Beers2005,Masseron2010,Placco2014} in the sense that these are evolved giants, they reside in binary systems, and the secondary is very likely a white dwarf (provided that HD 59686 B is confirmed to be a white dwarf). Recently, \citet{Jorissen2016a} has identified low-amplitude RV variations superimposed on the binary trend in 3 CEMP stars in a sample of 13. They show periods of $\text{about one}$ year and RV semi-amplitudes of hundreds of m/s. \citet{Jorissen2016b} discussed the origin of the RV variations of one system in particular, HE 0017+005, and suggested that this may be due to pulsations in the envelope of the giant star. Unfortunately, the spectral types of the stars from Jorissen et al. are not well established. The authors assumed that all the stars have masses of $\sim$0.9 \Msun, and it is likely that these very metal-poor stars are in a different stage of the stellar evolution than HD 59686 A, which we expect to be on the HB with a 89\% probability \citep[see][]{Reffert2015}. In particular, the \logg~values of the RV-variable CEMP stars seems to be much lower than that of HD~59686~A \citep[see][]{Jorissen2016b}, which makes pulsations much more plausible for those stars. Even if pulsations should be confirmed as the correct interpretation of the RV variations observed in CEMP stars, this will probably not be the case for HD 59686 A because all the available observational evidence at hand (e.g., H$\alpha$ index, photometry, infrared RVs) supports the planetary hypothesis, unless some exotic not-yet-observed surface macroturbulent structure or long-lived long-period non-radial pulsation was taking place in HD~59686~A for more than a decade, which we consider unlikely. Nevertheless, there is much that we do not know about long-period stellar oscillations in giant stars, and we cannot fully exclude such phenomena.

\subsection{Discarding a hierarchical triple star system}
Another possibility that can mimic planets in binary systems are hierarchical triple systems in which the observed RV signals are caused by another star orbiting the binary companion instead of a planet orbiting the primary star. For instance, \citet{Schneider2006} and \citet{Morais2008} studied the effects on the RV measurements of a star orbited by a pair in a close circular orbit in a triple star system. They concluded that the effect of the binary is approximately weaker than $\sim$1 m/s in the RV semi-amplitude and can only mimic a low-mass Earth- or Saturn-like planet. Later, \citet{Morais2011} extended their work to triple star systems on eccentric orbits, showing that the binary effect is stronger than in the cirular case. However, the magnitude of the RV semi-amplitude is still about a few meters per second and cannot account for the large variation that we see in the RV data of HD~59686~A ($K\sim$137 \ms). Furthermore, we can estimate the effect that a binary star system with a total mass of $\sim$0.5~\Msun~with a period of $\sim$300 days can generate in the RV semi-amplitude \citep[using Eq. 37 of][]{Morais2011}. For reasonable values of the amplitudes of the frequency terms induced by a hypothetical third star in the system, we obtained a value of the RV semi-amplitude mimicking a planet ranging from $\sim$1$-$5~\ms, that is, more than an order of magnitude smaller than what we observe in our data. We therefore conclude that a hidden star orbiting the stellar object HD~59686~B is not the cause of the observed RV variations in the system.

\section{High-contrast images}
\subsection{Previous search for stellar companions in HD~59686 A}
HD~59686 A has been examined before for stellar companions.
\citet{Roberts2011} found a visual component separated by 5.61\arcsec.  Assuming a face-on circular orbit, this corresponds to a minimum separation of $\sim$519~au. If this component were a physical companion, then the separation would lead to an orbital period far too large to be visible in our data. \citet{Baines2008b} have also observed HD~59686 A using the CHARA interferometer. They performed fits to the diameter of several stars and found that single stars were consistently fit with low values of  $\chi^2$, while the presence of a stellar companion created a systematic behavior in the residuals, resulting in a high $\chi^2$ value.  They saw no such systematic behavior in the fit of HD~59686 A and therefore ruled out a MS companion more massive than G5~V within a field of view of 230 mas ($\sim$23~au). 

\begin{figure*}[]
\resizebox{\hsize}{!}{\includegraphics{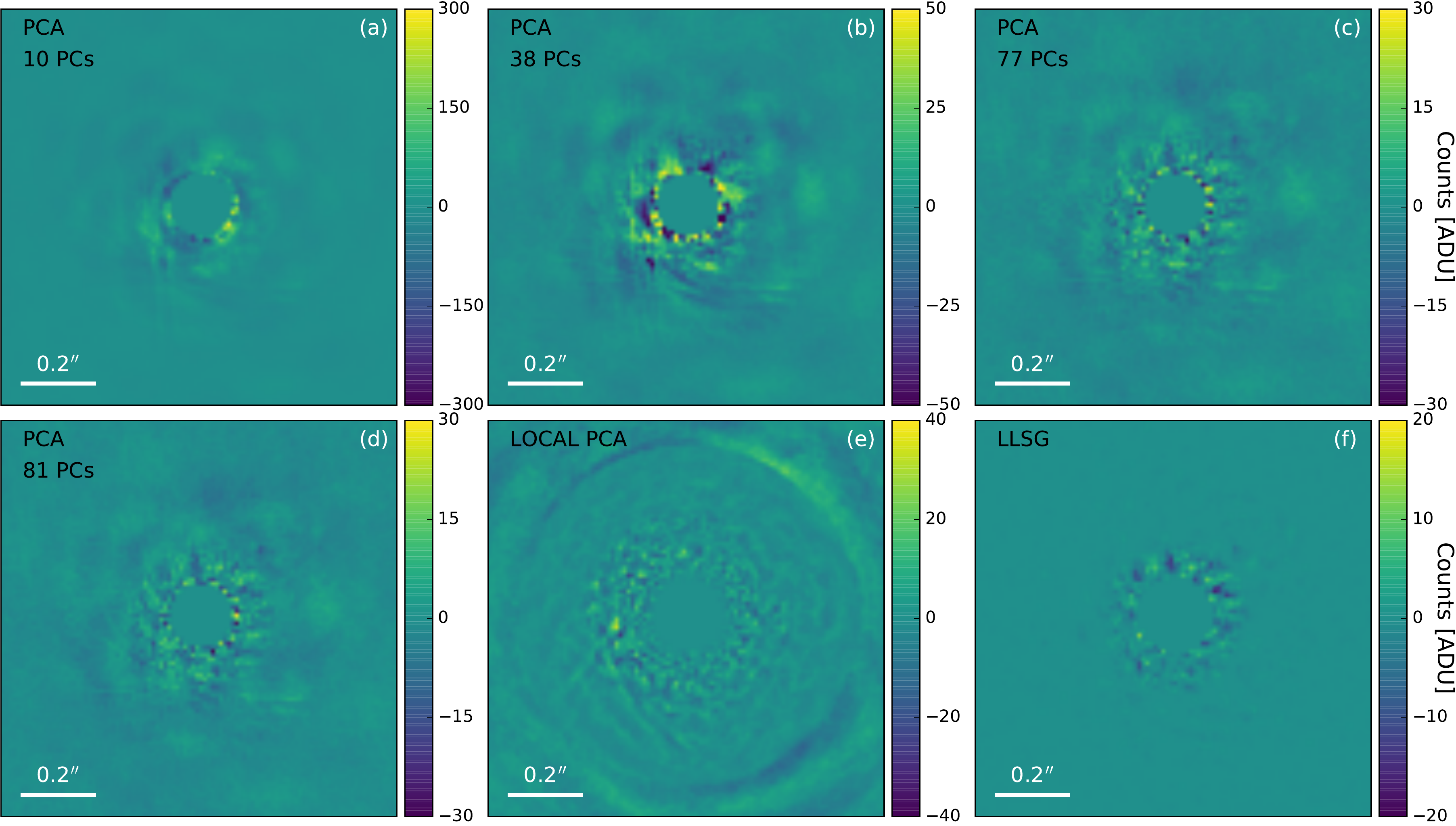}}
\caption{High-contrast $L'$-band LMIRCam images of HD~59686 A. Panels $a$, $b$, $c,$ and $d$ show the residual images after running the PCA with 10, 38, 77, and 81 principal components. Panel $e$ shows the image obtained with a local, subannular PCA approach, and panel $f$ presents the residual image after subtracting the stellar PSF using the new LLSG algorithm. No signal of the companion star HD~59686 B is detected in any of the panels.}
\label{lbt_images}
\end{figure*}

\citet{Baines2008b} also searched for small-separation binaries by looking for separated fringe packets in the data.  If a second star were present in the system with a separation of around 10 to 100~mas ($\sim$1--10~au), then two sets of fringe packets would be detected. However, no separated fringe packet was observed for HD~59686 A. This approach relies on the assumption that the angular separation of the two stars is not small and that the position angle is not perpendicular to the projected baseline angle. Most likely, the authors failed to detect HD~59686 B because this star is expected to be much fainter than the giant primary and was probably below the contrast sensitivity of CHARA.

With the aim of investigating the nature of the stellar object HD~59686 B, we acquired high-resolution images of this system as explained in the following sections.

\subsection{Observations and data reduction}
The high-contrast imaging observations of HD~59686 A were carried out on February 9, 2014 using the L/M-band InfraRed Camera \citep[LMIRCam;][]{Skrutskie2010,Leisenring2012} mounted at the Large Binocular Telescope Interferometer \citep[LBTI;][]{Hinz2012} on Mt. Graham, Arizona. LMIRCam is a high-resolution camera designed to operate in the 3$-$5~$\mu$m wavelength range. The infrared detector is a 1024$\times$1024 HgCdTe array, with a plate scale of 10.707$\pm$0.012 mas/pix \citep{Maire2015} and a field of view of 11$\times$11\arcsec. 

The observations were taken using only the left side of the LBT in pupil-stabilized mode, which further allows the use of angular differential imaging \citep[ADI;][]{Marois2006}. The core of the PSF was intentionally saturated to increase the signal of the binary companion. Unsaturated exposures with a neutral density filter were also taken for calibrating the photometry. The AO system was locked with 300 modes during the whole duration of our observations. We obtained 205 minutes of on-source integration and $\sim$100$^{\circ}$ of field rotation. A total of 7\,413 images of HD~59686 A were taken in the $L'$-~band filter ($\lambda_{c}$=3.70 $\mu$m, $\Delta\lambda$=0.58 $\mu$m). 

To properly subtract the background emission and detector systematics, the star was dithered to two different positions on the detector separated by 4.5$''$. Additionally, our reduction steps included dark current subtraction, flatfielding, bad pixel correction, bad image removal, image alignment, and trimming of the data. We were left with a 300$\times$300 pixel datacube of 5487 reduced images. However, during large parts of the observing sequence, weather conditions were not optimal (seeing $>$1.5\arcsec), so that we decided to discard 20\% of the images based on the measurement of the correlation of each one of the frames with respect to a high-quality reference frame of the sequence. In total, we obtained a datacube of 4389 images. 

\subsection{PSF subtraction}
In addition to simple ADI processing, more sophisticated algorithms
exist, such as the locally optimized combination of images \citep[LOCI;][]{Lafreniere2007} and principal component analysis \citep[PCA;][]{Amara2012,Soummer2012,Brandt2013}.
They can be used to subtract the light profile of a star to detect possible companions around it. We decided to follow a PCA approach, as it has been shown to produce better contrast performance for small inner working angles \citep[e.g.,][]{Meshkat2014}. The expected binary separation at the time of our observations is small, so that even with the PCA technique it is challenging to detect any signal at all, considering that we do not know the orbital inclination and orientation of the orbit. 

To analyze our stack of images, we used the open-source Python package VIP\footnote{\href{https://github.com/vortex-exoplanet/VIP}{\color{blue}https://github.com/vortex-exoplanet/VIP}} \citep{Gomez2016a}, which provides a collection of routines for high-contrast imaging processing, including PCA and slight variations of it, such as annular and subannular PCA. The PCA algorithm models the star light as a linear combination of a set of orthogonal basis functions or principal components (PCs) and fits for the PC coefficients in each of the frames in the stack. This means
that the parameter that must be set is the number of PCs used to model the PSF in each frame. We started by estimating the optimal number of PCs by inserting a star in each of the images at a small separation from the center of the primary star. We varied the magnitude difference of this fake companion with the central star from  $\Delta m$=8$-$11 mag in steps of 0.5 mag and determined the number of PCs that maximizes the S/N in an aperture of 1 FWHM centered on the coordinates of the fake star after running the PCA. We searched in a grid ranging from 1-200 PCs and found that the highest S/N values were obtained for 10, 38, 77, and 81 PCs. The central saturated core of the PSF (eight-pixel radius) was masked and not considered in the fitting. We show in Fig. \ref{lbt_images}, panels $a$ to $d,$  the results after running the PCA in the stack of images of HD~59686 A using the previously derived numbers of PCs. No significant signal was found in the residual images.

Additionally, we also performed a local PCA by fitting for the stellar PSF in quadrants of circular annuli of 3 FWHM width around the central star. In this case, the PCA is computed locally in each quadrant, and we applied a parallactic angle rejection of 1 FWHM to discard adjacent frames and avoid self-subtraction of the companion star. The number of PCs was decided automatically in each quadrant by minimizing the residuals after subtracting the PSF. The resulting resdiudals image is shown in panel $e$ of Fig. \ref{lbt_images}. As in the full-frame PCA, no significant companion is seen in the plot.

\begin{figure}[t]
\resizebox{\hsize}{!}{\includegraphics{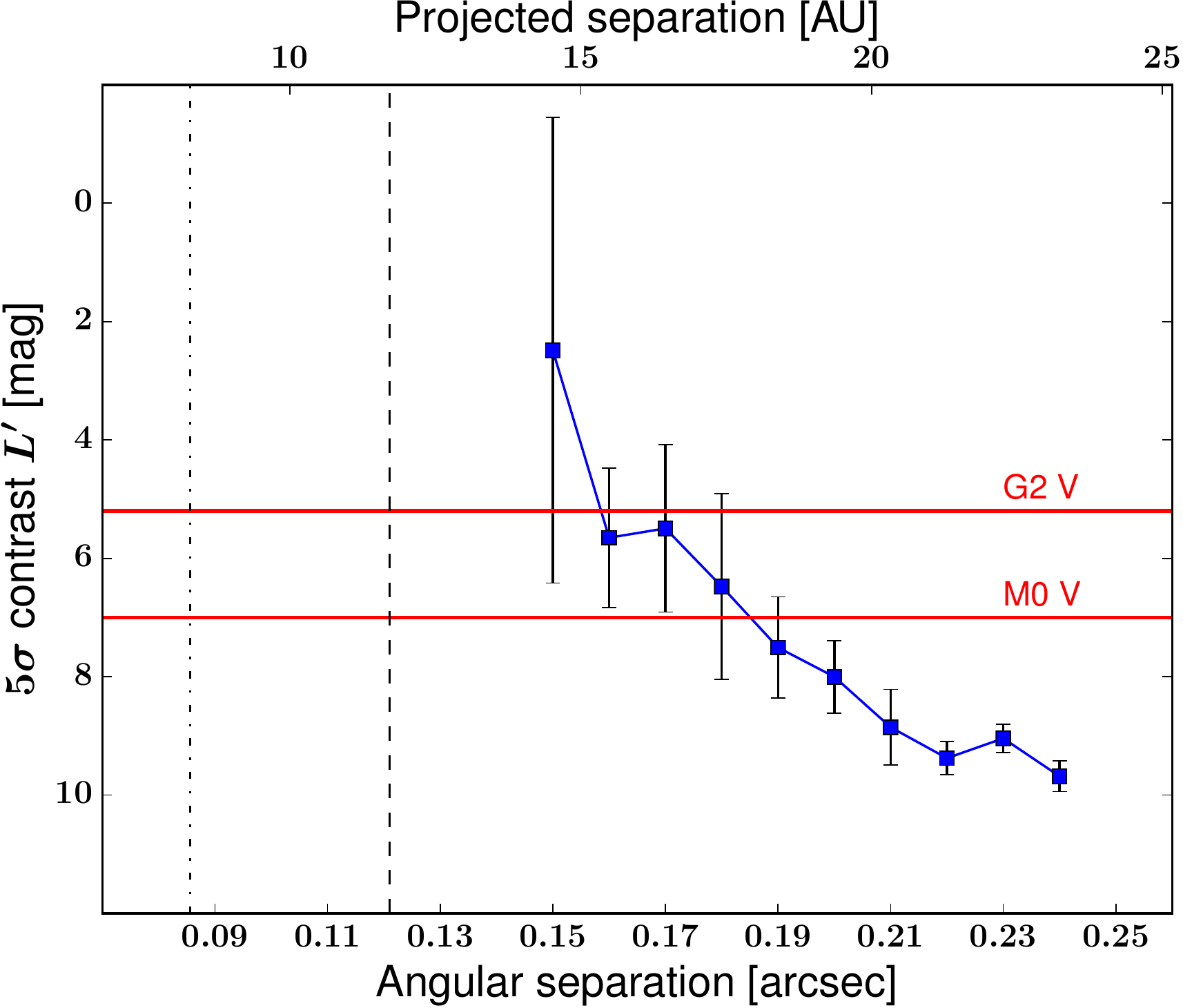}}
\caption{5-$\sigma$ detection limits in terms of the magnitude contrast in the $L'$-band as a function of the distance from the central star. The black dashed line represents the binary separation upper limit of $\sim$11.7~au at the time of the observations. The dash-dotted line marks the saturation radius of $\sim$0.085\arcsec. The red solid lines mark the expected contrasts for a G2 V star of 1 \Msun~and a  M0 V star of 0.5 \Msun, from top to bottom.}
\label{contrast_limit}
\end{figure}

As an alternative to the standard PCA, we also used the new algorithm recently introduced by \citet{Gomez2016b} to subtract the stellar PSF of high-contrast images and enhance the signal of faint companions. The method is named by the authors local low-rank plus sparse plus Gaussian-noise decomposition (LLSG). The main idea of the algorithm is to use a robust PCA approach \citep[see, e.g.,][]{Candes2009} to decompose the stellar image into three components; a low-rank, a sparse, and a Gaussian noise part. The low-rank carries most of the signal from the stellar PSF, the Gaussian noise captures the quasi-static features of the speckle noise, and the sparse component contains the signal of potential faint companions. The most important parameter to set in the LLSG algorithm  is the rank, which is equivalent to set the number of PCs in the standard PCA. We chose a rank of 51 as the mean of the optimum number of PCs derived before. We note, however, that varying the rank number does not change the obtained results significantly. The residual image after the LLSG subtraction is shown in panel $f$ of Fig. \ref{lbt_images}. Although the quality of the image seems to be much better than in previous images, we did not detect any signal from the binary star HD~59686 B.  

The obtained results can be explained by (i) the poor weather during some part of the observations, (ii) the small expected angular separation of the companion, and (iii) the probability that the orbit orientation placed the star at a projected separation such that the companion is not visible from Earth at the time of observation. 

\subsection{Contrast curve calculation}
Assuming that the orbital configuration is favorable at the observing time and that we are only limited by the contrast of the binary pair and the quality of our images, we can set constraints on the maximum brightness that the companion star could have without being detected in our images. To do that, we injected fake companions of various magnitudes at different distances from the central star. As a fake companion star we used the median-combined PSF of the unsaturated data set and scaled it to different contrast ratios based on the photometry of the unsaturated image of HD~59686 A and taking into account the different exposure times between the saturated and unsaturated frames. The fake stars were then inserted in each of the reduced stack of images, accounting for the change in parallactic angle during the rotation sequence. We then processed these images with the VIP package in the exact same way as before and calculated the 5-$\sigma$ detection limit in terms of S/N at the position of each fake star. We adopted the S/N defintion of \citet{Mawet2014} as we are working at distances very close to the center of the star, and the low-pixel statistics applies. We repeated this procedure at four different position angles for each radius and then took the average to minimize random speckle errors.

In Fig. \ref{contrast_limit} we show the 5$\sigma$ contrast curve of the LBT images as a function of the angular separation from the central star. Our data reach contrasts between $\sim$5--10 mag for separations between $\sim$0.16\arcsec--0.24\arcsec (15.5--23.2~au). We also show the maximum expected binary separation at the observing time of $a_{max}\sim11.7$~au (black dashed line) and the PSF saturation radius of $r_{s}\sim8.3$~au (black dash-dotted line). The expected separation of the binary pair comes from a detailed study of the dynamical stability of the HD~59686 system that constrains the orbital inclination to the range $i\sim50^{\circ}$--90$^{\circ}$ (Trifonov et al. 2016, in prep.). For an inclination of $50^{\circ}$, we derived the value of $11.7$~au, which translates into $\sim$0.12\arcsec of angular separation. Adopting higher values for the inclination results in lower values for the binary separation. Unfortunately, the large saturation radius of the LMIRCam images ($\sim$0.085\arcsec) prevents us from deriving reliable values for the 5$\sigma$ contrast in the region $\lesssim0.15$\arcsec~($\lesssim14.5$ au), in which we expect HD~59686 B to reside. Nevertheless, we show in the plot (red solid lines) the expected contrasts for a star of 0.5 and 1~\Msun. A G-type star of 1~\Msun~or greater is excluded for separations $\gtrsim17$ au. For lower masses and separations our sensitivity decreases significantly, and we cannot exclude the presence of a star with masses between 0.5$-$1~\Msun.

\begin{figure}[t]
\resizebox{\hsize}{!}{\includegraphics{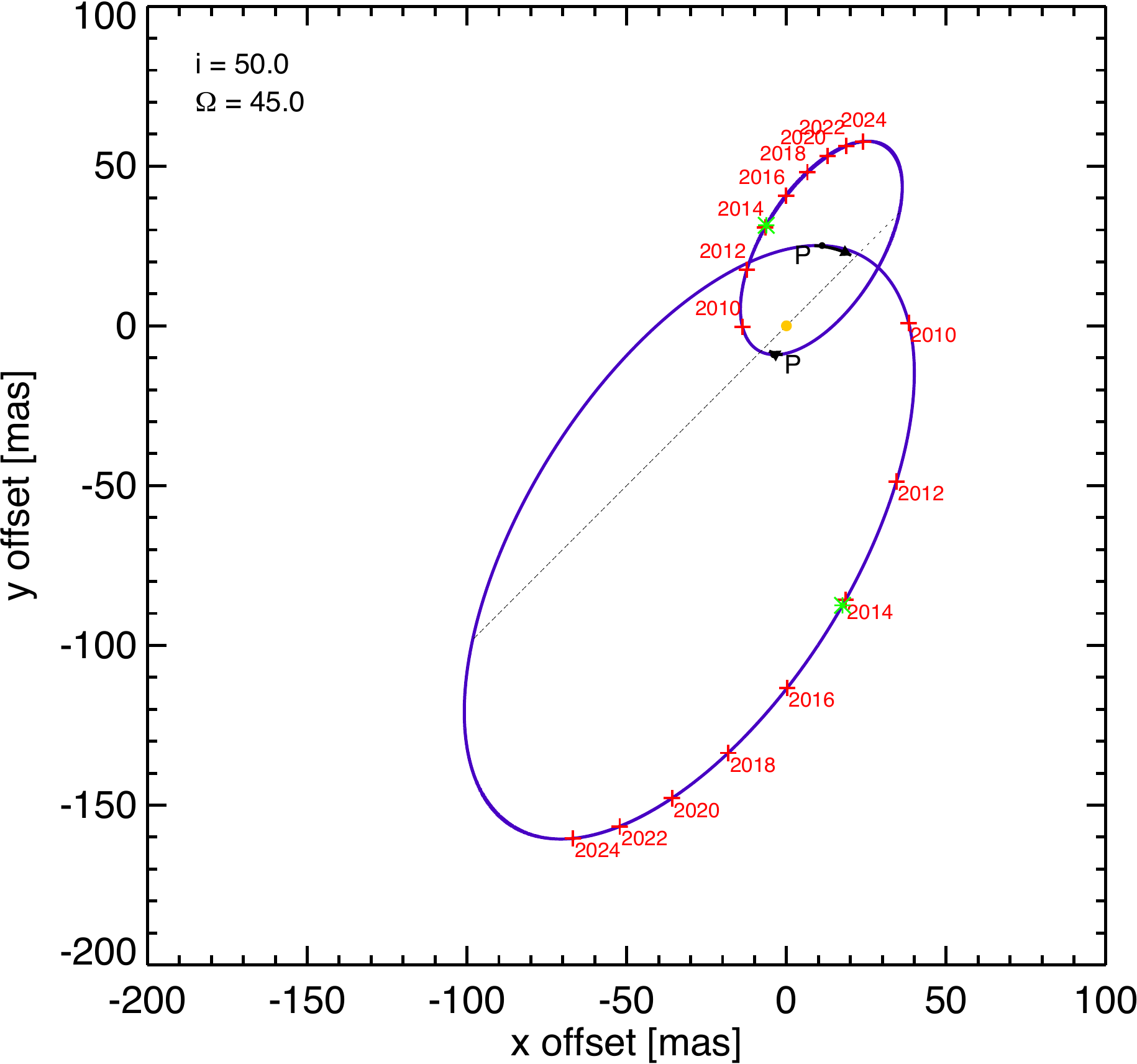}}
\caption{Sky-projected orbit of the HD~59686 binary system assuming values of $i=50^{\circ}$ and $\Omega=45^{\circ}$ for the orbital inclination and longitude of the ascending node, respectively. Labeled in the orbit are the positions of each star as a function of time. The green symbol marks the position of HD~59686 A and HD~59686 B at the time of our LBT observations. The dotted line is the line of nodes, and the letter P denotes the positions of the stars at periastron. The yellow dot marks the center of mass of the system.}
\label{binar_orbit}
\end{figure}

To illustrate the configuration of the binary system, we show in Fig. \ref{binar_orbit} the sky-projected orbit of HD~59686~AB derived from the fitted orbital parameters. The red labels mark the position of each of the stars at certain times (in years). The green symbols highlight the respective locations of HD~59686~A and HD~59686~B in the binary orbit at the time of the LBT observations. The high eccentricity of the binary is clearly visible. Fortunately, both components are moving away from each other at the moment, so that it should become easier to detect HD~59686~B in the coming years. In about $\sim$2025, the system will be in apastron at a minimum separation of roughly $\sim$20--21~au assuming an inclination of $i=90^{\circ}$. For lower values of the inclination the binary separation increases. Future high-resolution observations of this system are highly encouraged to better constrain the nature of the stellar object HD~59686~B. 


\section{Discussion}
\subsection{HD~59686 Ab: a planet in a close-separation binary}
\begin{figure}
\resizebox{\hsize}{!}{\includegraphics{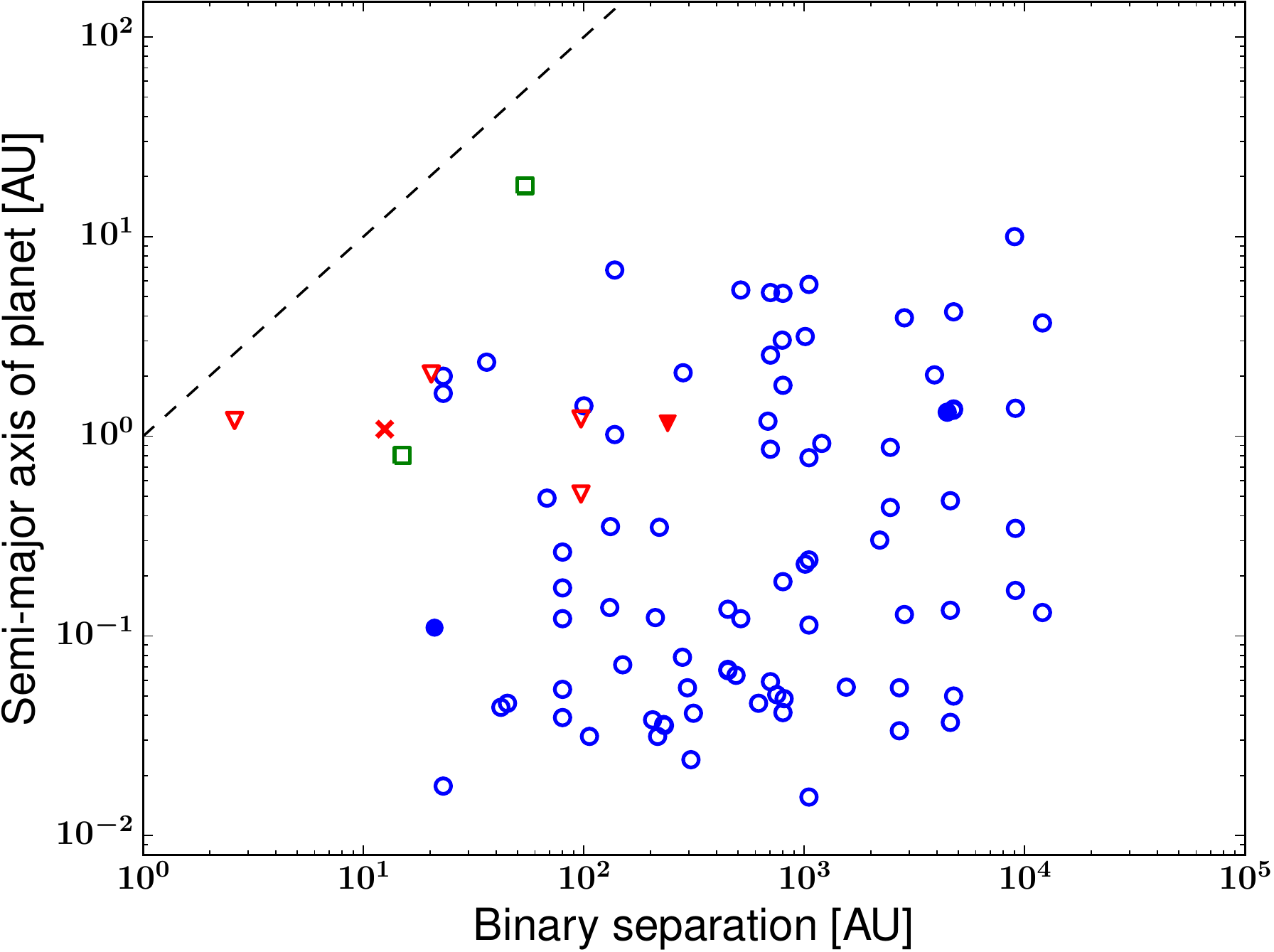}}
\caption{Semi-major axis of planetary companions plotted against binary
  separation for all known planet-hosting binary systems. Shown are binaries with MS (blue circles) and evolved subgiant/giant (red triangles) primary stars as well as two microlensing binaries (green squares), in which the spectral type of the stars is not known. The filled symbols show binaries in which the secondary star is a white dwarf. The position of the HD~59686 system is marked with a red cross. The dashed line marks the 1:1 relation between planet semi-major axis and binary separation. Most of the discovered planets are found in binary stars with separations greater than $\sim$100~au.}
\label{binaxis}
\end{figure}

Among the known S-type planets, HD~59686 Ab is very peculiar, mainly because it is part of a close-separation ($a_{B}=13.6$~au) and eccentric ($e_{B}=0.7$) binary system. Figure \ref{binaxis} shows the semi-major axis of the known S-type planets as a function of the binary separation. Planets exist in binaries with a wide range of separations, but it is clear that the majority of them show semi-major axes greater than $a_{B}\sim100$~au. HD~59686 AB is, together with $\nu$ Octantis \citep{Ramm2009} and OGLE-2013-BLG-0341LB \citep{Gould2014}, the binary with the closest separation of its stellar components known to harbor a planet. 


The microlensing Earth-mass planet OGLE-2013-BLG-0341LB b is orbiting at approximately $\sim$0.8 au from its host star, and the microlensing models are compatible with a binary separation of either $\sim10$ or $15$ au. The case of $\nu$ Oct is particularly remarkable, since the separation of the binary pair is only $a_{B}\sim2.6$~au and the conjectured planet is orbiting at $a_{p}\sim1.2$~au; roughly at half the distance between both stars. Interestingly, similar to HD~59686 AB, the $\nu$ Oct system is composed of a single-lined K-giant binary, with a secondary star mass of $\sim0.55$~\Msun. Moreover, the $\nu$ Oct system is slighlty eccentric: $e\sim0.25$ \citep{Ramm2015}. As we discussed below, the existence of giant planets in both systems is very hard to explain by traditional theories.

There are two additional systems (not included in the plot) with reported companions at $a\lesssim20$~au: KOI-1257 \citep{Santerne2014} and $\alpha$ Cen \citep{Dumusque2012}. KOI-1257 b is a transiting giant planet with a period of $P=86.6$ days that is part of a binary system with $a_{B}\sim5.3$~au. However, the nature of the massive outer companion in the system is unconstrained at present; it could be anything, a planet, a brown dwarf or a stellar object \citep{Santerne2014}. On the other hand, in $\alpha$ Cen AB, the stellar nature of the binary components is well established, but the existence of a terrestial planet orbiting at $\sim0.04$~au has recently been questioned \citep{Hatzes2013,Rajpaul2016}, implying that most likely there is no planet in the $\alpha$ Cen system. This would make HD~59686 AB,  and $\nu$ Oct, unique systems in which to study the formation of giant planets in short-separation binaries. 


\begin{figure*}
\begin{multicols}{2}
\includegraphics[width=\linewidth]{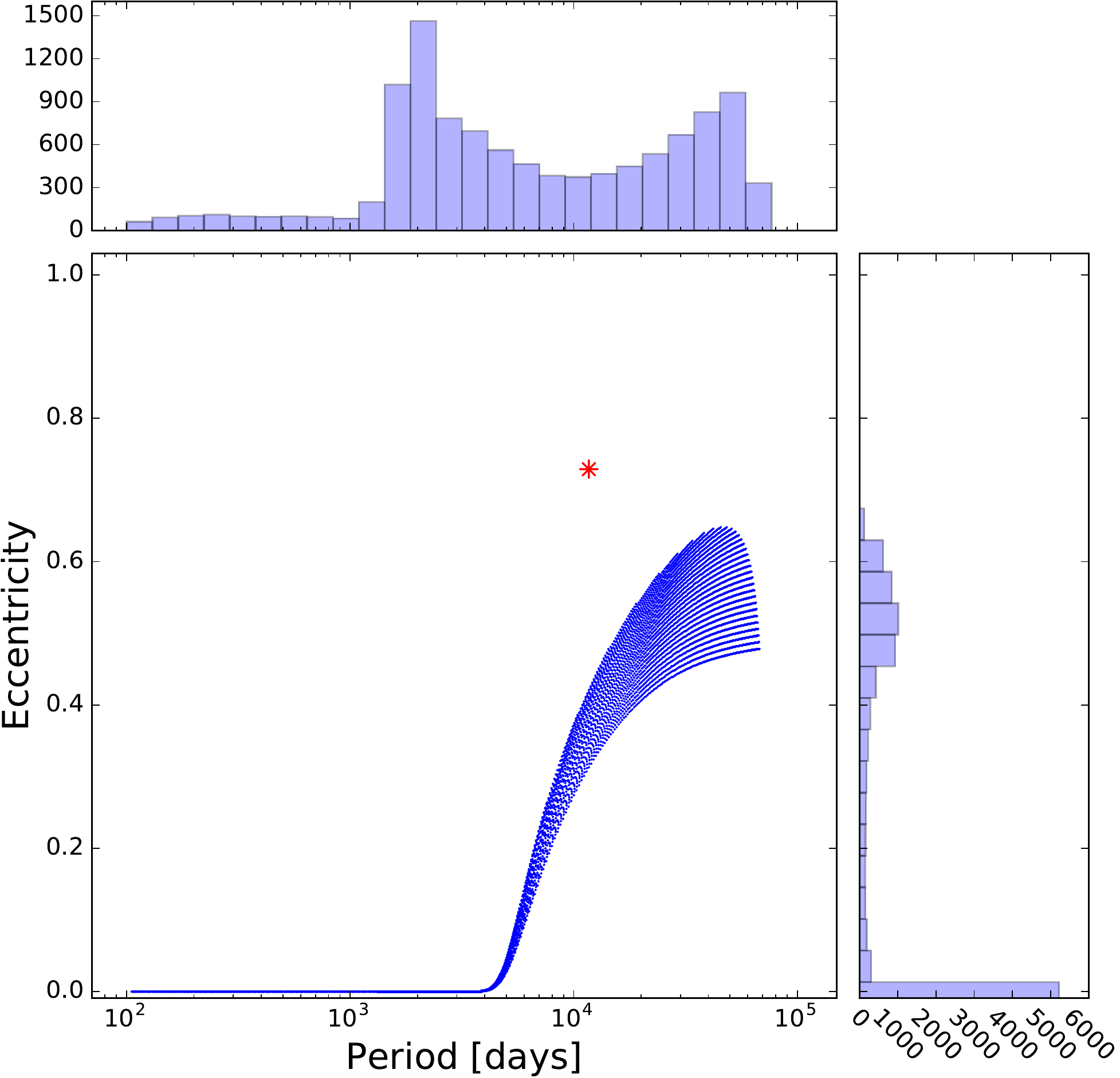}\par
\includegraphics[width=\linewidth]{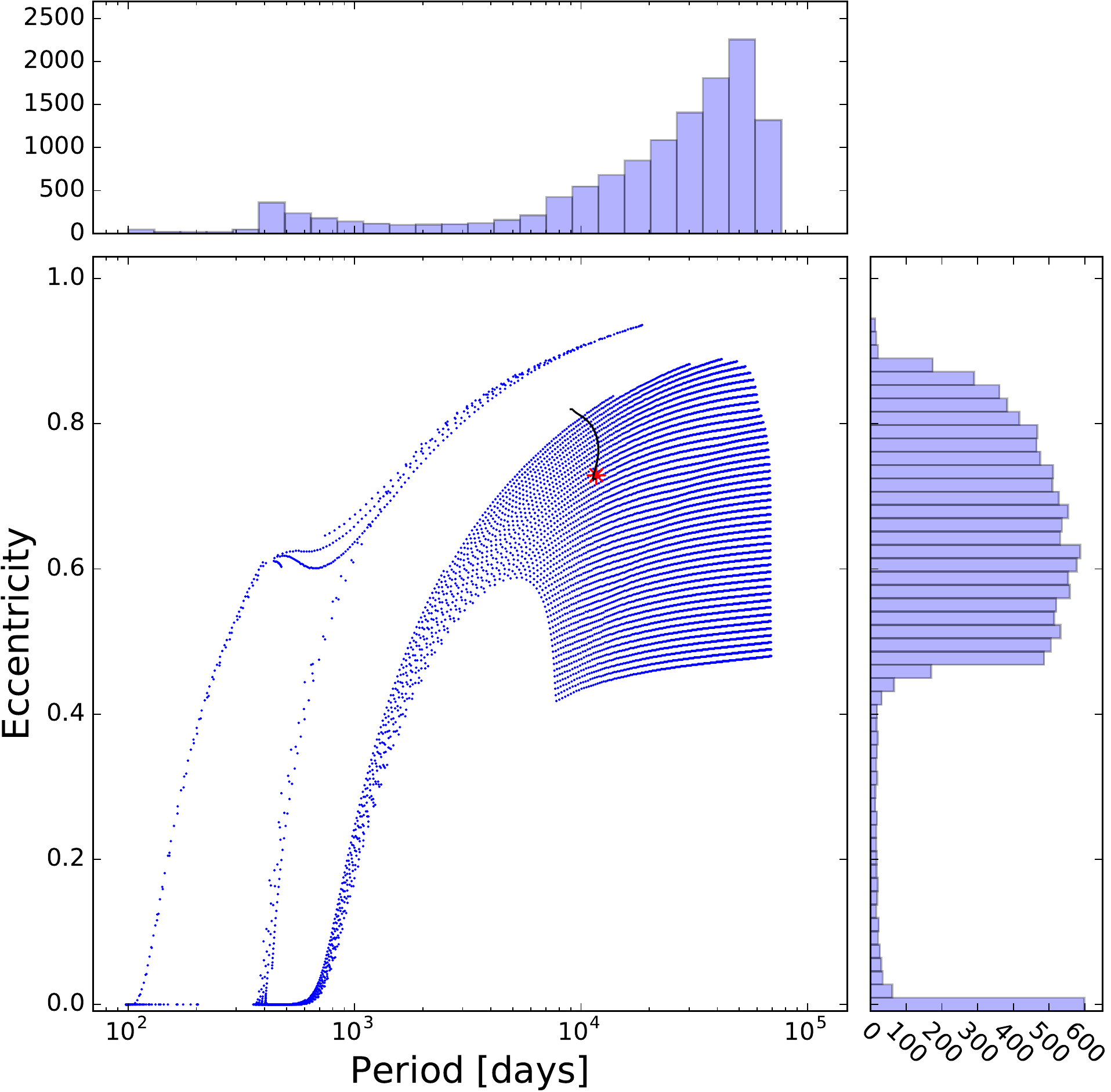}\par
\end{multicols}
\caption{Final periods and eccentricities resulting from all simulations that led to the formation of a HB giant star with a WD companion. The initial masses of the stars were 1.9 and 2.3 \Msun. We also show histograms reflecting the distribution of final periods and eccentricities. The position of the HD~59686 system is marked with a red asterisk. \textit{Left panel}: Results for $B_{W}=0$, meaning that the mass loss is treated with the traditional Reimers prescription. It is clear that none of the simulations can reproduce the HD~59686 system. In the majority of the cases the orbit is fully circularized due to tidal interactions during the AGB phase. \textit{Right panel}: Results for an enhanced mass-loss rate ($B_{W}=10^{4}$) showing that a large fraction of initial orbital conditions lead to long-period and eccentric binaries similar to the HD~59686 system. The black solid line marks the period and eccentricity evolution of the model that agrees best with the orbital properties of HD~59686 AB.}
\label{bse_code}
\end{figure*}

Another striking property of the HD~59686 system is the high eccentricity of the binary pair. With a value of $e_{B}=0.729_{-0.003}^{+0.004}$, this is the most eccentric close-separation binary ($a_{B}\lesssim20$~au) known to harbor a planet. This implies that, at periastron, both stars are separated by only $\sim$3.6~au. The formation of such a system presents a tremendous challenge to current planet formation theories as the smallest binary separation in which giant planets could form is thought to be $\sim20$~au \citep[see][and references therein]{Haghighipour2009}. On the other hand, simulations have shown that terrestrial planets may form in close-separation binaries up to $\sim0.2q_{b}$, where $q_{b}$ is the binary pericenter distance \citep{Quintana2007}. Therefore, this possibility is not directly excluded in the HD~59686 system, as terrestial planets might have formed up to a distance of $\sim$0.7~au from the primary star. 

\subsection{Nature of the stellar object HD~59686 B}
With the mass for HD~59686 B constrained in the range 0.53--0.69~\Msun~ derived from dynamical simulations, there are two options for this stellar companion: it may be a typical dwarf star or, more interestingly, a white dwarf (WD). The latter possibility is not rare as there are currently three known circumstellar planets orbiting stars with WD companions: GL 86 \citep{Queloz2000}, HD 27442 \citep{Butler2001}, and HD 146513 \citep{Mayor2004}. Interestingly, the system GL 86 AB is also a close-separation binary with a semi-major axis of $a_{B}=18.4$~au. With the currently available data we cannot assess the nature of the stellar object HD~59686 B with certainty, but nevertheless, we can investigate whether the WD scenario is plausible given the current orbital parameters and derived masses of the HD~59686 system.

If HD~59686 B is indeed a WD, then its mass must originally have been greater than the mass of HD~59686 A (1.9 \Msun) because it evolved faster to a later stage of the stellar evolution. The problem now resides in estimating the inital MS mass of HD~59686~B. The initial-final mass relationship (IFMR) for WDs has been a subject of intense research in the past \citep{Weidemann1977,Weidemann1987,Weidemann1990,Jeffries1997}. More recently, \citet{Kalirai2009} calibrated a semi-empirical relation for the IFMR using several WDs found in a set of globular clusters in the Milky Way. With this relation we can estimate an inital MS mass for HD~59686 B of $\sim$0.7$-$2.3 \Msun~for WD masses of 0.53 and 0.69 \Msun, respectively. The latter mass satisfies our intial constraint of $M_{B}>1.9$ \Msun. This means that, for the upper limit of our mass estimate, HD~59686~B may have evolved off the MS to end its life as a WD of $\sim$0.69 \Msun.

To investigate whether this scenario is plausible, we used the detailed binary evolution code BSE \citep{Hurley2002} to evolve a binary star pair with a set of initial orbital properties. The initial binary masses were set to 2.3 and 1.9 \Msun. We considered a range of initial periods and eccentricities of $P=5000-30000$ days in steps of 100 days and $e=0.50-0.99$ in steps of 0.01. The system was then evolved until the stellar types of the two stars were a WD and a HB star. The results of the simulations are shown in Fig. \ref{bse_code}, where the final periods and eccentricities are plotted for all the different initial orbital configurations that led to a WD-HB pair with similar masses as those observed in the HD~59686 system. The left panel shows the results for a mass-loss prescription given by the traditional Reimers formula for red giants \citep{Reimers1975}. It is clear that no set of initial conditions can reproduce the current orbital properties of the HD~59686 system, namely a period of $P=11680$ days and eccentricity of $e=0.729$. Orbits with periods of a few thousands days or less are fully circularized, and the small fraction of systems with a high eccentricity ($e\sim0.6$) shows very long orbital periods of $\sim$50000 days.

The right panel of Fig. \ref{bse_code} shows the final periods and eccentricities for the same initial configurations as discussed before, but with an increased mass-loss rate controlled by the enhanced wind parameter $B_{W}$. This parameter was first introduced by  \citet{Tout1988} to explain the mass inversion of the RS CV binary star Z Her. In this scenario it is assumed that the mass loss is enhanced through tidal interactions with a binary companion. Tidally enhanced stellar winds have been used since then to account for several phenomena related to giant stars in binary systems, such as the eccentricities of barium stars \citep{Karakas2000,Bonacic2008}, symbiotic channel for SNe Ia progenitors \citep{Chen2011}, morphology of HB stars in globular clusters \citep{Lei2013}, and long-period eccentric binaries with He WD \citep{Siess2014} and SdB companions \citep{Vos2015}. 

The efficiency of the tidally enhanced stellar wind was set to $B_{W}=10^{4}$ by \citet{Tout1988} to fit the observed parameters of Z Her, but this value may vary depending on the specific system considered. The results plotted in the right panel of Fig. \ref{bse_code} are for a value of $B_{W}=10^{4}$, but we note that we are able to reproduce the orbital parameters of the HD~59686 system with several values of $B_{W}$ ranging from $\sim$5000$-$10000. A striking difference with the case of a standard mass loss is that now a considerable fraction of the simulations shows eccentric orbits in the range $\sim$0.40$-$0.85 with periods of a few tens
of thousand days, very similar to HD~59686 AB. The model that best reproduces HD~59686 AB (black solid line) has a final eccentricity and period of $e_{f}=0.724$ and $P_{f}=11555$ days (with initial values of $e_{i}=0.82$ and $P_{i}=9000$~days), very close to the actual observed values of the HD~59686 system. These results show that the WD scenario for HD~59686 B is plausible, provided that its progenitor passed through an enhanced wind mass-loss phase during the AGB evolution. It is worth mentioning, however, that the previous calculations do not include the presence of a planet in the binary system. If the planet HD~59686 Ab existed before the presumed evolution of the stellar companion HD~59686 B, then the change from MS star to giant star to white dwarf could have affected the evolution of the planetary body. 

Regardless of the nature of the stellar object HD~59686 B, the formation of a planet with a stellar companion at 13.6 au with a periastron distance of only 3.6 au presents serious challenges to standard planet formation theories \citep[e.g.,][]{Hatzes2005}. In the core-accretion model \citep[e.g.,][]{Mizuno1980,Pollack1984,Lissauer1993}, giant planets close to their host stars are expected to form beyond the snow line and then migrate inward to reach their current positions. For a mass of $\sim$1.9~\Msun~the snow line of HD~59686 A is located at $\sim$9.7~au \citep[assuming the model of][]{Ida2005}. However, with an eccentric stellar companion at 13.6 au, the protoplanetary disk around the primary star would be truncated at around 1 au or less \citep{Pichardo2005}, preventing the formation of a giant planet at this separation from the host star \citep{Eggenberger2004}. Similarly, a formation in situ at $\sim1-2$~au by disk instability \citep[e.g.,][]{Kuiper1951,Toomre1964,Boss2000} is highly unlikely as the required temperature for efficient cooling would be too high for the protoplanetary disk to remain bound to the central star \citep{Rafikov2005}. Additionally, giant planets are not expected to form by disk instability in binary systems with separations of $a_{B}\lesssim20$~au and eccentricities of $e_{B}\gtrsim0.4$ \citep{JangCondell2015}.

\subsection{Possible origin of the planet HD~59686 Ab}

With the increasing number of planets found in non-conventional configurations in binary systems, in both P-type and S-type orbits,  new mechanisms have been proposed to explain their origin. For instance, \citet{Schleicher2014} developed a model to explain circumbinary planets in the close binary NN Ser from the ejecta of common envelopes. They also extended their model to predict the masses of 12 planetary candidates around post-common envelope binaries (PCEBs) listed by \citet{Zorotovic2013}, showing a good agreement in several systems. Additionally, \citet{Perets2010} and \citet{Tutukov2012} have discussed the possibility of forming second-generation (SG) circumstellar planets in evolved binary systems. The main idea of SG planets is that an evolved star transfers mass to its companion, and when the binary separation is small enough, this could lead to the formation of an accretion disk around the primary star with sufficient mass to form planets. If the stellar object HD~59686 B is confirmed to be a WD, then this scenario appears as an interesting alternative to explain the origin of HD~59686~Ab. 

In principle, this system would satisfy several expected observational characteristics from SG planets. As stated by \citet{Perets2010}, SG planets are expected to be almost exclusively found in evolved binary systems with compact objects, such as WD or neutron stars. They are also likely to be more massive than normal first-generation planets; with a mass roughly constrained between $\sim$7$-$9~\Mjup, HD~59686 Ab is among the most massive exoplanets detected so far. SG planets could reside in regions of orbital phase space forbidden to pre-existing planets by dynamical arguments. HD~59686 Ab is marked as unstable or on the border of stability by some dynamical criteria \citep{Holman1999,Mardling2001}, although detailed N-body integrations allow stability for a certain parameter space including both prograde and retrograde orbital configurations (Trifonov et al. 2016, in prep.). For the prograde case, the bootstrap dynamical test yielded a small sample of long-term stable fits consistent with the bootstrap distributions at the 1 sigma confidence level. These prograde fits are locked in secular resonance with aligned orbital periapsis. The best dynamical fits assuming a retrograde orbit have slightly better quality (smaller $\chi^{2}$) and are long-term stable. It is worth noting that there is evidence suggesting that the planet in the $\nu$ Oct system, that is, the tight binary with $a_{B}=2.6$~au and a K-giant primary,  is in a retrograde orbit \citep{Eberle2010,Gozdziewski2012,Ramm2015}. 

Although the SG planet scenario may seem attractive, we cannot discard the possibility that the current configuration of the HD~59686 system may be the result of past dynamical interactions in the native star cluster \citep{Pfahl2006}. In this context, the planet HD~59686~Ab could have formed beyond the snow line around its single host star, and later, through dynamical processes, another binary star may have exchanged one of its stellar members for this single star with the already formed planet. This scenario has been invoked in the past to explain the origin of a giant planet in the system HD 188753 \citep{Pfahl2005,Zwart2005}. However, the existence of this planet was recently proved false by \citet{Eggenberger2007}. \citet{Pfahl2006} estimated that dynamical interaction in the parent star clusters would deposit giant planets in roughly $0.1$\% of binary systems with semi-major axis of $a<50$~au. We note that this value was obtained under several assumptions and it is unlikely that we have detected such a system in our sample, which does not contain a large number of such binaries.

Another similar, albeit slightly different possibility is that the present configuration of the HD~59686 system might have been generated in the past after the formation of the planet HD~59686~Ab was completed. In this scenario, planets can form in wide-separation binary systems that are not hostile for the planet formation process and later, through a close stellar encounter or a perturbation induced by a former third star in the system, the orbital parameters of the system may have changed to those observed today. This possibility was first suggested by \citet{Marzari2007}, who studied the dynamical evolution of triple star systems with a primary star harboring a planet. They found that close stellar encounters or a perturbation of the original triple system may significantly change the binary orbit, leading to more eccentric and tight binaries with planets. Additionally, using numerical simulations, \citet{Marti2012} studied the formation of the planet around $\gamma$ Cep by stellar scattering and found that around $\sim$1$-$5\% of fly-by encounters involving planetary systems could lead to planets in close-separation binaries. Although this number is small, we cannot exclude this possibility for the formation of HD~59686 Ab. 

\section{Conclusions}
By obtaining high-precision RVs of the giant star HD~59686~A for more than 12 yr, we discovered a clear RV signature most likely caused by a massive ($m_{p}~\sini=6.92_{-0.24}^{+0.18}$~\Mjup) giant planet, HD~59686~Ab, at a distance of $a_{p}=1.0860_{-0.0007}^{+0.0006}$~au from its host star. Additionally, we detected the strong signal of an eccentric ($e_{B}=0.729_{-0.003}^{+0.004}$) binary companion, HD~59686~B, orbiting with a semi-major axis of only $a_{B}=13.56_{-0.14}^{+0.18}$~au. This makes HD~59686~AB, together with $\nu$ Oct,  the binary system with the closest separation of its stellar components known to harbor a giant planet. Furthermore, at periastron, the
two stars are separated by just $3.6$~au; a certainly hostile environment for the formation of any planet in this system.

We acquired high-resolution images of HD~59686~A using LMIRCam at the LBT telescope with the aim of investigating the nature of the stellar object HD~59686~B. We could not directly detect the star, mainly because the small expected angular separation ($\lesssim0.12$\arcsec) from the host star poses great challenges to current PSF-subtraction techniques. It is most likely that the binary companion is a red dwarf star or a white dwarf. The binary system will be at apastron in about 2025, with an expected separation of the binary pair of around $\sim$20--21~au. With a favorable orbital configuration it would be possible to detect the companion with a similar strategy as we followed in this work.

Regardless of the nature of the binary companion, the existence of a planet in an eccentric binary with a separation of $\lesssim15$~au is a challenge for standard planet formation theories, namely core accretion and disk instability. It is expected that massive giant planets form in massive protoplanetary disks with $M_{d}\gtrsim10^{-2}$~\Msun. In the HD~59686 system, a disk would be tidally truncated at roughly $\sim$1 au \citep{Pichardo2005}, resulting in a disk not massive enough for the formation of giant planets \citep{JangCondell2015}. Additionally, stirring by the tidal field may inhibit the growth of icy grains and planetesimals and also stabilize the disk against fragmentation \citep{Nelson2000,Thebault2004,Thebault2006}. Under these conditions, the in situ formation by disk instability is not a plausible mechanism for giant planet formation. However, \citet{Rafikov2015} have recently shown that it is possible to form planets within $\sim$20 au separation binaries, provided that the protoplanetary disks are massive and only weakly eccentric. It would be interesting to test the validity of this model in the HD~59686 system.

As a different approach to the origin of HD~59686 Ab, we discussed the possibility that this planet could have formed in a second-generation protoplanetary disk, assuming that the stellar object HD~59686 B is a white dwarf. We demonstrated that given the current properties of the system, this scenario is feasible, and discussed its implications regarding the formation of HD~59686 Ab. Altough not directly verifiable with the currently available data, the second-generation planet hypothesis is an attractive alternative for the origin of HD~59686 Ab as this system accounts for several observational characteristics for this type of planets \citep[see][]{Perets2010}. Another mechanism that may explain the origin of the planet, although unlikely and hardly verifiable, is the past exchange of stellar companions through dynamical interaction with the native star cluster.

Our detailed analysis of the extensive RV data set of HD~59686~A supports the hypothesis that planets can exist in close binary 
systems with separations of $a_{B}\lesssim20$~au, contrary to the theoretical expectations \citep{Whitmire1998,Nelson2000} and the recent observational support showing that short-separation binaries are rarely found among \textit{Kepler} planet hosts \citep{Wang2014}. However, the question of how such planets may have formed remains unanswered as none of the standard theories can satisfactorily explain the origin of HD~59686~Ab. In this context, systems such as HD~59686  and $\nu$ Oct  may become benchmark objects in the study of planet formation in tight binaries.  

\begin{acknowledgements}

We would like to thank the staff at the Lick Observatory for their support during
the years of this project. We are also thankful to the CAT observers who
assisted with this project, including Saskia Hekker, Simon Albrecht, David
Bauer, Christoph Bergmann, Stanley Browne, Kelsey Clubb, Dennis K{\"u}gler,
Christian Schwab, Julian St{\"u}rmer, Kirsten Vincke, and Dominika
Wylezalek. We also thank the staff of the LBT for carrying out the high-contrast image observations of this system. The LBT is an international collaboration among institutions in the United States, Italy and Germany. LBT Corporation partners are: The University of Arizona on behalf of the Arizona university system; Istituto Nazionale di Astrofisica, Italy; LBT Beteiligungsgesellschaft, Germany, representing the Max-Planck Society, the Astrophysical Institute Potsdam, and Heidelberg University; The Ohio State University, and The Research Corporation, on behalf of The University of Notre Dame, University of Minnesota and University of Virginia. This research has made use of the SIMBAD database and the VizieR catalog access tool, CDS, Strasbourg, France.  This publication made use of data products from the Two Micron All Sky Survey, which is a joint project of the University of Massachusetts and the Infrared Processing and Analysis Center/California Institute of Technology, funded by the National Aeronautics and Space Administration and the National Science Foundation.
This research made use of Astropy, a community-developed core Python package for Astronomy \citep[][\href{http://www.astropy.org}{\color{blue}http://www.astropy.org}]{Astropy2013}. This work used the python package astroML \citep[][\href{https://github.com/astroML/astroML}{\color{blue}https://github.com/astroML/astroML}]{Vanderplas2012} for the calculation of the GLS periodogram. The plots in this publication were produced using Matplotlib \citep[][\href{http://www.matplotlib.org}{\color{blue}http://www.matplotlib.org}]{Hunter2007}. T. T. and M. H. L. were supported in part by the Hong Kong RGC grant HKU 17305015. M.O. thanks V. Bailey for useful discussions regarding the high-contrast images taken with LMIRCam. We are grateful to the anonymous referee for providing insightful arguments regarding alternative interpretations of RV variations.

\end{acknowledgements}



\end{document}